\documentclass[aps,superscriptaddress,showpacs,reprint,floatfix]{revtex4-1}

\usepackage{graphicx}
\usepackage{color}
\usepackage[colorlinks=true, urlcolor=black, linkcolor=black, citecolor=black]{hyperref}
\usepackage{float}
\usepackage{caption}
\usepackage{subcaption}
\usepackage{physics}
\usepackage{url}

\begin{document}

\title{Incidence of the Brownian relaxation process on the magnetic properties of ferrofluids}

\author{Lili Vajtai}
\affiliation{{Department of Physics, Institute of Physics, and HUN-REN-BME Condensed Matter Research Group, Budapest University of Technology and Economics, M\H{u}egyetem rkp. 3., H-1111 Budapest, Hungary}}

\author{Ferenc Simon}
\email{simon.ferenc@ttk.bme.hu}

\affiliation{{Department of Physics, Institute of Physics, and HUN-REN-BME Condensed Matter Research Group, Budapest University of Technology and Economics, M\H{u}egyetem rkp. 3., H-1111 Budapest, Hungary}}
\affiliation{{Institute for Solid State Physics and Optics, HUN-REN Wigner Research Centre for Physics, PO. Box 49, H-1525, Hungary}}

\author{Maria del Puerto Morales}
\affiliation{Department of Nanoscience and Nanotechnology, Instituto de Ciencia de Materiales de Madrid (ICMM-CSIC), 28049 Madrid, Spain}

\author{Kolos Molnár}
\affiliation{{Department of Polymer Engineering, Faculty of Mechanical Engineering, Budapest University of Technology and Economics, M\H{u}egyetem rkp. 3., H-1111 Budapest, Hungary}}
\affiliation{{HUN–REN–BME Research Group for Composite Science and Technology, M\H{u}egyetem rkp. 3., H-1111, Budapest, Hungary}}
\affiliation{{MTA-BME Lendület Sustainable Polymers Research Group, M\H{u}egyetem rkp. 3., H-1111 Budapest, Hungary}}

\author{Balázs Gábor Pinke}
\affiliation{{Department of Polymer Engineering, Faculty of Mechanical Engineering, Budapest University of Technology and Economics, M\H{u}egyetem rkp. 3., H-1111 Budapest, Hungary}}



\author{Norbert Marcel Nemes}
\affiliation{Departamento de Física de Materiales, Universidad Complutense de Madrid,
28040 Madrid, Spain}

\begin{abstract}

Ferrofluids containing magnetic nanoparticles represent a special class of magnetic materials due to the added freedom of particle tumbling in the fluids. We studied this process, known as Brownian relaxation, and its effect on the magnetic properties of ferrofluids with controlled magnetite nanoparticle sizes. For small nanoparticles (below 10 nm diameter) the Néel process is expected to dominate the magnetic response, whereas for larger particles, Brownian relaxation becomes important. Temperature- and magnetic field-dependent magnetization studies, differential scanning calorimetry, and AC susceptibility measurements were carried out for 6, 8, 10.6, and 13.5 nm diameter magnetite nanoparticles suspended in water. We identify clear fingerprints of the Brownian relaxation for the sample of the large diameter nanoparticles as both magnetic and thermal hysteresis develop at the water freezing temperature, whereas the samples of small diameter nanoparticles remain hysteresis-free down to the magnetic blocking temperature. This is supported by the temperature-dependent AC susceptibility measurements: above 273 K, the data show a low-frequency Debye peak, which is characteristic of the Brownian relaxation. This peak vanishes below 273 K. 
\end{abstract}

\maketitle

\section{Introduction}

Nanomagnetic ferrofluids remain the focus of interest due to several unexpected phenomena as well as some intriguing applications such as lubricants, medical imaging contrast materials \cite{mri_ferrof}, or magnetothermal heating substances in medical therapy \cite{app1,appl2,ferrohyd}. Although the fundamental magnetic properties of ferrofluids have been studied \cite{alapkut1,alapkut2,alapkut_puerto,alapkut4}, their high-frequency magnetic response is less explored \cite{highf1,highf2}. Knowledge of these is crucial for magnetic hyperthermia applications as it sets the optimal working conditions (e.g. frequency) for radiofrequency irradiation to avoid overlap with unwanted side effects (e.g. heating of the body caused by eddy currents) \cite{HERGT}. 

The frequency-dependent power absorption in nanomagnetic fluids is governed by the imaginary part of the dynamic susceptibility (or $\chi "$). In its simplest mono-exponential form, this results in a Debye relaxation behavior, characterized by a single relaxation time, $\tau$. This, in turn, is governed by two different relaxation processes. In the first one, the particle is static and its magnetization rotates (known as the Néel process) and in the second one, the entire particle rotates (known as the Brownian process) while its magnetization is unchanged with respect to the particle \cite{ferrohyd2}. The first one is typical for smaller particles (typically with diameters below 10 nm, this value is dependent both on the atomic structure and the geometry of the nanoparticles) and the second one occurs for larger particles \cite{tau_d}. 

Temperature- and magnetic field-dependent studies in nanomagnetic ferrofluids focused mostly on Néel dominated samples \cite{alapkut4,neel1,neel2,neel3,neel4,neel5,neel6} using DC magnetometry, with a limited effort for larger particles with Brownian relaxation \cite{alapkut_puerto,alapkut4,jackpot}. These effects were also theoretically studied for various models including ferrohydrodynamics \cite{ferrohyd,ferrohyd2}, the Fokker–Planck equation, or the egg model using the Landau--Lifsitz--Gilbert theory \cite{FP_egg,elm2,elm3}. The most common approach in the investigation of dilute ferrofluids is to use linear response theory \cite{relax} but it is also possible to consider higher-order terms in the magnetic response \cite{elm3}. Several studies focused on the preparation and characterization of a particular nanomagnetic ferrofluid using magnetometry \cite{jackpot,synth1,synth2,synth3,langevin}.
AC susceptibility was also studied using both frequency- and time-domain approaches \cite{highf1,highf2,tdom} but no information exists on the temperature- and frequency-dependent contributions of the two processes, albeit it is highly relevant for clinical applications in nanomagnetic hyperthermia \cite{Review_ortega_pankhurst,Review_Garaio,Review_Wu}. 

Here, we study water-based ferrofluids containing magnetite (Fe$_3$O$_4$) nanoparticles of well-controlled and distinctively different diameters. The nominal average diameters were 6, 8, 10.6, and 13.5 nm; the Néel relaxation dominates the 6-nm sample, whereas the Brownian relaxation is expected to be the dominant process in the 13.5-nm sample. 
We performed temperature-dependent DC magnetometry, including field-cooled (FC) and zero-field-cooled (ZFC) measurements, and identified the main differences in the two sets of results caused by different relaxation mechanisms being present. The experiments were accompanied by DSC (differential scanning calorimetry) measurements to precisely monitor the freezing of water and to correlate its effect with the magnetometry results. Temperature-dependent AC susceptibility measurements clearly identify the contributions of the two types of processes. The presence of a characteristic, temperature-dependent magnetic hysteresis (taken at low magnetic fields) around the solvent freezing temperature is identified as a straightforward fingerprint for Brownian relaxation.

\section{Methods}

The magnetic nanoparticles were obtained following the co-precipitation protocol described by Massart \cite{Massart_np_preparation}, by introducing small modifications to control particle size. The particle size was measured with transmission electron microscopy (TEM), resulting in 5.8$\pm$1.5 nm for the 6-nm sample and 13.2$\pm$4 nm for the 13.5-nm sample. Most reported data were taken on sufficiently diluted samples with 3 mg/mL concentration, although we studied the concentration dependence up to 30 mg/mL where some influence was detected and the results are discussed in the Supporting Information.

A Quantum Design MPMS3 device was used in DC scan mode in the static magnetometry measurements. Typically 10-20 $\mu$L of the investigated ferrofluids were permanently sealed with a torch in quartz tubes. The magnetic field offset due to the history-dependent trapped superconducting flux was corrected by comparing to hysteresis cycles of a paramagnetic Pd reference.


To study the so-called magnetic blocking effect, we performed zero-field-cooled (ZFC) and field-cooled (FC) experiments; for the latter, we used 10 kOe. We employ SI units throughout the manuscript except for the magnetometer output in CGS units. The temperature ramp rate was set to 2 K/min and the calorimetric studies validated that this protocol results in the proper thermalization of the solutions.

 Differential scanning calorimetry (DSC) measurements were performed with a TA Instruments DSC Q2000 device. Sample holders with unfixed lids fabricated for this instrument were filled with 10-20 $\mathrm{\mu L}$ of the investigated ferrofluid suspensions. During measurements, the samples were uniformly cooled and heated between $\mathrm{25~^\circ C}$ and $\mathrm{-50~^\circ C}$, and the required heat flow was recorded. Multiple measurement cycles were performed with both the heating and the cooling rate being 1, 2, 5, 10, and 20 K/min. Below 5 K/min (speed of both cooling and heating), results were no longer affected by further slowing down the rate. This shows that the 2 K/min ramping rate in the magnetometry results in proper thermalization. Peaks in the heat flow data are signs of first-order phase transitions, such as the freezing and the melting of the carrier fluid.

A small amount of carrier fluid may evaporate during lengthy measurements (a typical value being 30 minutes) but repeated cycling revealed that it did not affect the DSC results; neither the heat flow nor the the observed transition temperatures. 

Frequency-dependent magnetic susceptibility was measured using the AC measurement system (ACMS) option of the Quantum Design PPMS system, between 10 Hz and 10 kHz using various excitation field amplitudes (0.1 Oe--10 Oe) in zero DC magnetic field. The AC susceptibility was not affected by the excitation field amplitude.

\section{Results and discussion}

\begin{figure}[h!]
    \centering
    \begin{subfigure}{\linewidth}
        \centering
        \includegraphics[width=\linewidth]{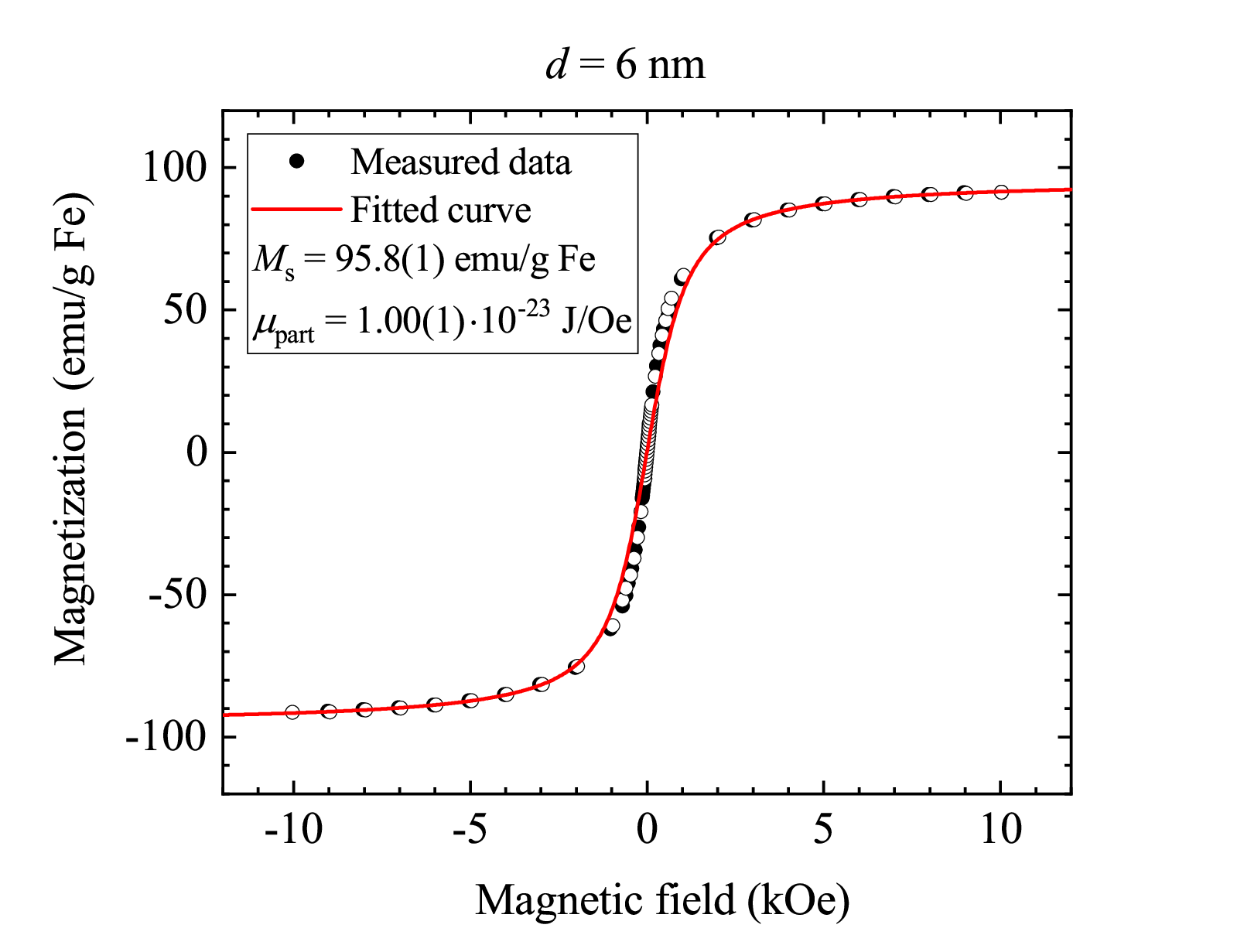} 
   \end{subfigure}
   \hfill
    \begin{subfigure}{\linewidth}
        \centering
        \includegraphics[width=\linewidth]{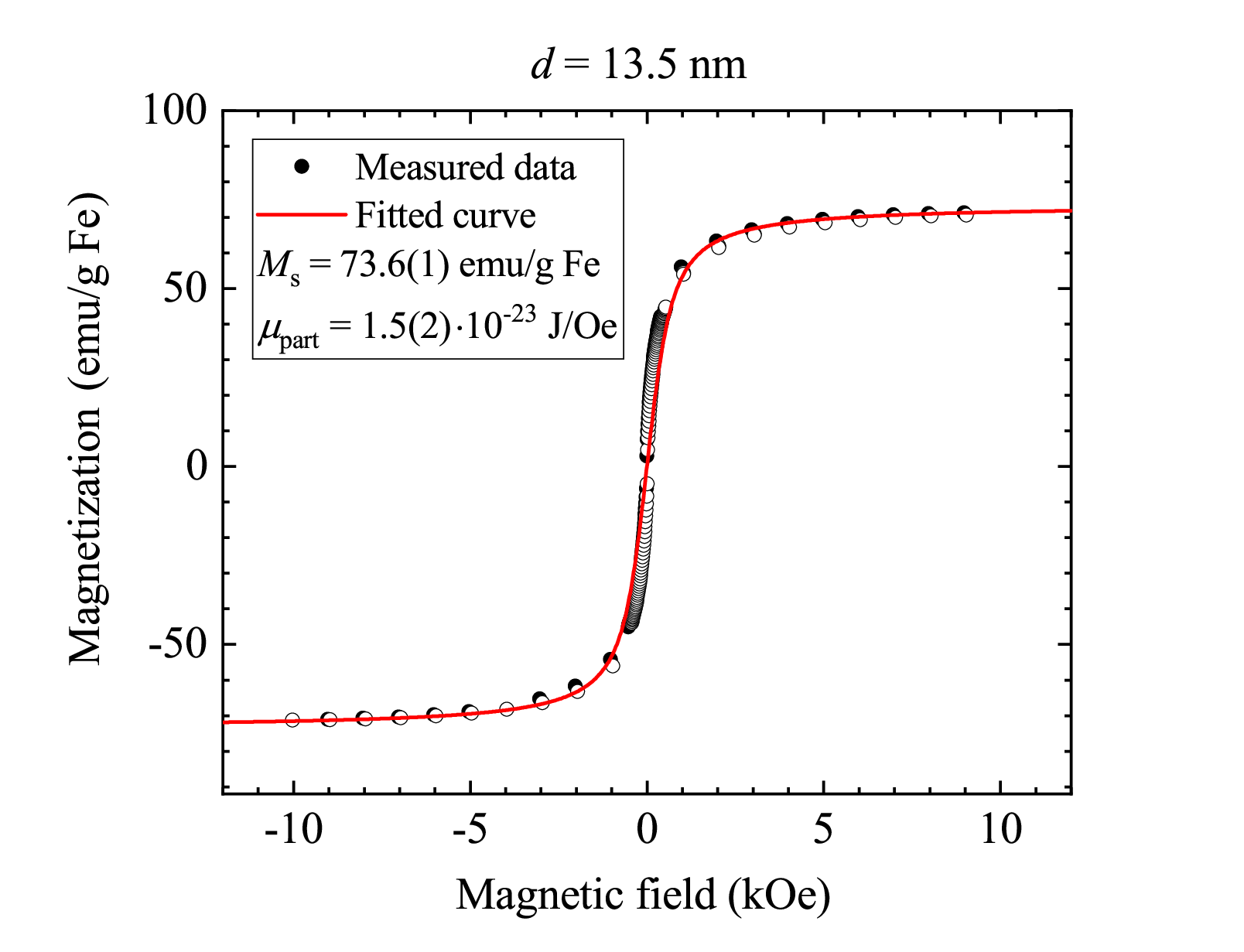}
        \end{subfigure}
    \caption{Magnetization curves for the two types of magnetite samples recorded at 300 K: containing 6-nm particles (top) and 13.5-nm particles (bottom). The fitted Langevin functions are shown with solid curves and fitting parameters. Full circles denote data recorded at decreasing magnetic field, whereas data with open symbols were taken at increasing magnetic field.}
    \label{lang}
\end{figure}

We primarily discuss nanoparticles with nominal diameters of $d=6$ nm and $d=13.5$ nm as these show the largest contrast between the Néel and Brownian relaxation behaviors. We also discuss intermediate diameter samples in the Supporting Information. Figure \ref{lang} shows the magnetization curves for the two types of samples at room temperature (magnetization is normalized by the mass of iron contained in the sample). Saturation occurs in a relatively low magnetic field (at a few 100 Oe), which indicates superparamagnetic behavior, the number of correlated spins being around 100-1000. The magnetic saturation of superparamagnetic nanoparticles is expected to follow a Langevin function (the high-spin limiting case of the Brillouin function) \cite{langevin}: 

\begin{equation}
\mathrm{L}(x) = \coth(x)-\frac{1}{x},
\end{equation}
which describes the sample magnetization as $M=M_{\text{s}}\mathrm{L}(x)$. where $M_{\text{s}}$ is the saturated magnetization and $x=\frac{H\mu_\text{part}}{k_{\text{B}}T}$. Here $k_\text{B}$ is the Boltzmann constant, $T=300~\mathrm{K}$ is the temperature, and $\mu_\text{part}$ is the magnetic moment of the superparamagnetic particle.

Figure \ref{lang} shows good Langevin fits the measured magnetization, yielding reasonable saturation magnetization values (95.8 emu/g Fe for the 6-nm sample and 73.6 emu/g Fe for the 13.5-nm sample). Literature values for $M_{\text{s}}$ are between 80 emu/g Fe$_3$O$_4$ \cite{Magnetite_80emu_per_g} and 86 emu/g Fe$_3$O$_4$ \cite{Magnetite_86emu_per_g}, which can be converted to 111 emu/g Fe and 119 emu/g Fe. The deviation of the measured data from these nominal values may stem from some uncertainty in the measurement of the nanoparticle amount in the suspensions and also from a slightly altered oxidation state of Fe ions on the nanoparticle surfaces. The latter is known to reduce the magnetization when freshly prepared magnetite transforms to maghemite with time \cite{magnetite_to_maghemite}. 

One expects that $\mu_\text{part}$ is about 8-10 times higher for the 13.5-nm sample than for the 6-nm sample (based on the volume ratio of the two particles) which is clearly not the case in the Langevin fits as quoted in Figure \ref{lang}. The Langevin fits do not account for any variation in nanoparticle diameter but are more sensitive for the value of $M_{\text{s}}$ and the result can be refined by inspecting the low-field data. In fact, we observe a deviation between the data and the fits in Figure \ref{lang}, which becomes more apparent when the numerical derivative of the data and the fits are inspected (data shown in the Supporting Information). This also allows us to obtain $\mu_\text{part}$ from the low-field derivative values. Given that the Langevin function can be approximated as 

\begin{equation}
\mathrm{L}(x) = \frac{x}{3}-\frac{x^3}{45}+\mathcal{O}(x^5)
\end{equation}
near the origin, the numerical derivative of the data at zero field also gives an estimate for $\mu_\text{part}$ as:
\begin{equation}
\left.\frac{\text{d}M}{\text{d}H}\right\vert_{H=0} = \frac{M_{\text{s}}\mu_\text{part}}{3 k_\text{B}T}
\end{equation}

Using the numerical derivative values of $0.14$ emu/g Oe  (6-nm sample) and $0.68$ emu/g Oe (13.5-nm sample), we obtain $\mu_\text{part}=1.82\cdot10^{-23}$ J/Oe (6-nm sample) and $\mu_\text{part}=1.14\cdot10^{-22}$ J/Oe (13.5-nm sample). Clearly, both values are larger than their counterparts from the Langevin fits, the change being almost a factor of 8 for the 13.5-nm sample. This difference between the permeability based on the saturation magnetization and the low-field slope is probably related to a distribution in the particle size (see Supporting Information). Alternatively, aggregation of the small grains may also influence the result. This affects the saturation and the permeability differently: saturation magnetization is independent of the particle size, whereas the low-field permeability depends linearly on the number of correlated spins in a superparamagnetic particle.

These refined $\mu_\text{part}$ values allow us to determine the number of correlated spins and the size of the nanoparticles given that the magnetic moment per Fe$_3$O$_4$ unit is 4.1 $\mu_{\text{B}}$ in magnetite \cite{kittel}. We obtain $d= 8.8$ nm (6-nm sample) and $d = 16$ nm (13.5-nm sample) and the number of correlated spins $N = 4.8\cdot 10^3$ (6-nm sample) and $N = 3\cdot 10^4$ (13.5-nm sample). These diameter values match reasonably well with those determined using TEM microscopy, especially since the magnetic data-based analysis may be affected by several factors, including inaccuracies in the concentration and mass measurements and a non-magnetic or a reduced-magnetic shell on the surfaces of the nanoparticles \cite{magnetite_to_maghemite}.  

\begin{figure}[h!]
    \centering
    \begin{subfigure}{\linewidth}
        \centering
        \includegraphics[width=\linewidth]{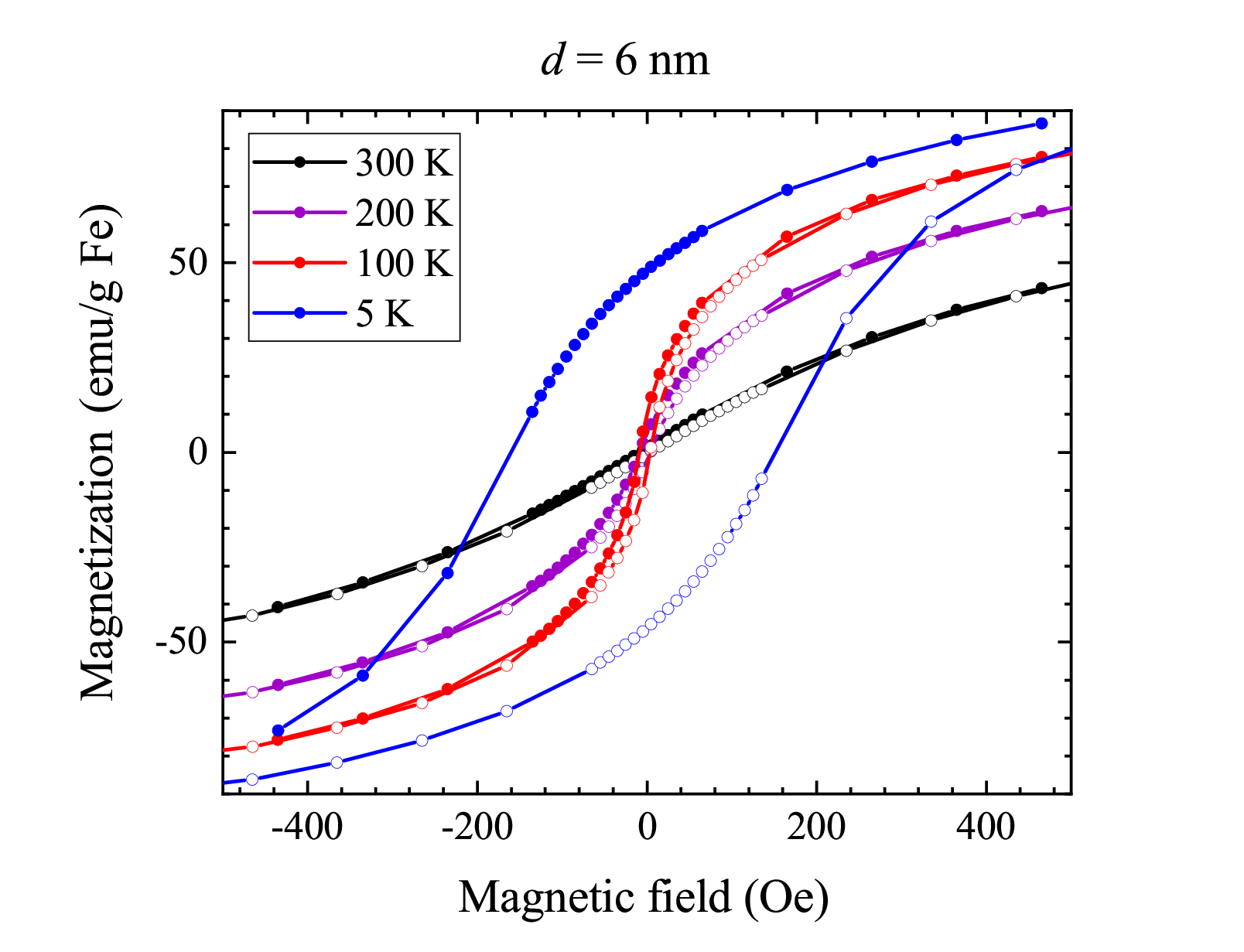} 
   \end{subfigure}
   \hfill
    \begin{subfigure}{\linewidth}
        \centering
        \includegraphics[width=\linewidth]{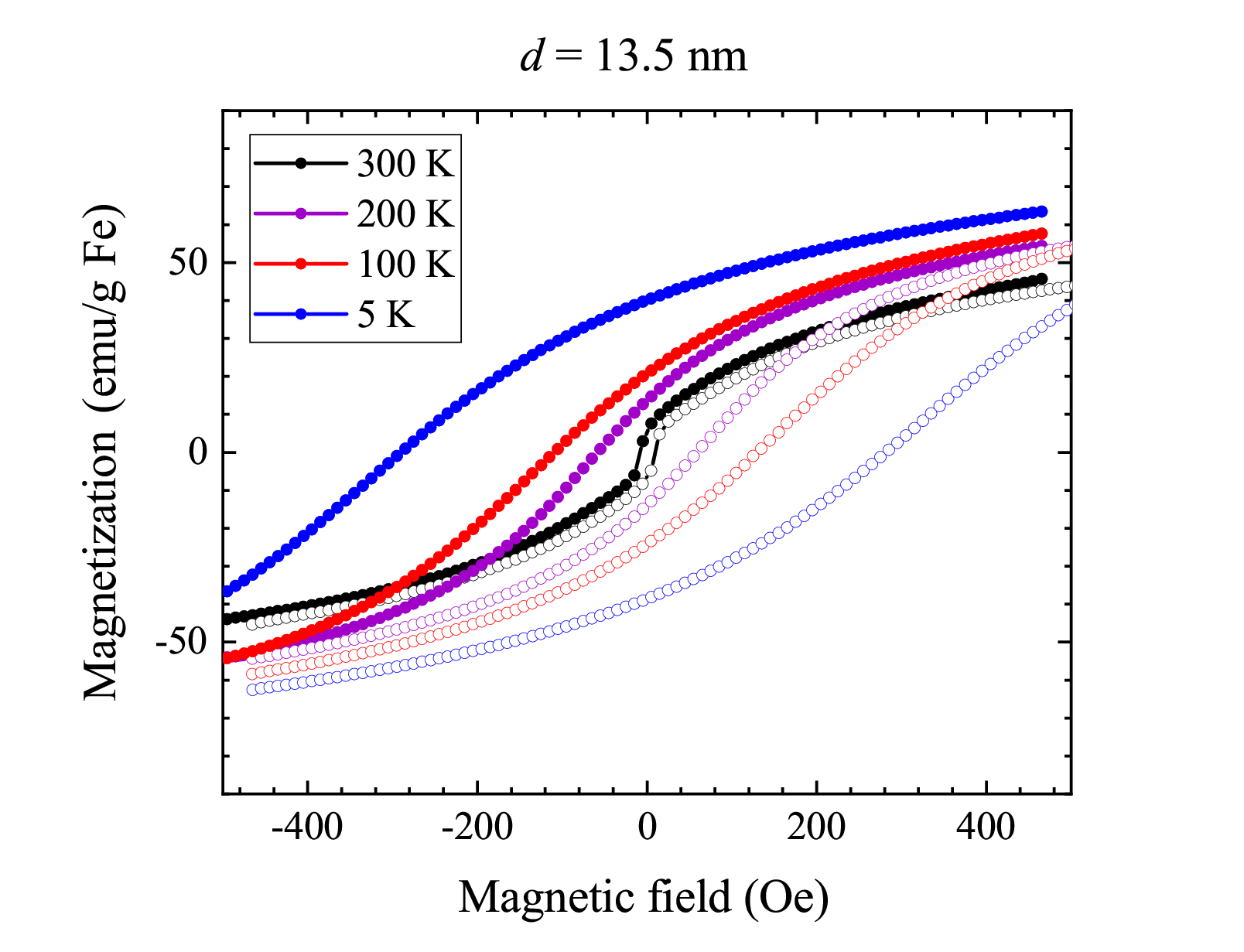}
        \end{subfigure}
    \caption{Magnetization hysteresis loops at different temperatures for the 6-nm sample (top) and the 13.5-nm sample (bottom). Note that hysteresis develops only below 100 K for the 6-nm sample, whereas it appears immediately below 300 K for the 13.5-nm sample. Full symbols denote data recorded at decreasing magnetic field, while data with open symbols were measured at increasing magnetic field.}
    \label{hist}
\end{figure}

Figure \ref{hist} shows the magnetic hysteresis curves at various temperatures for both samples. For the 6-nm sample, the coercive field is zero at room temperature with hysteresis appearing only below 100 K. This means that this sample remains superparamagnetic down to 100 K as expected for small-sized nanoparticles. In contrast, the 13.5-nm sample displays a hysteresis below 273 K, with the coercive field becoming progressively larger upon cooling. This observation shows that the freezing of water also hinders the nanoparticle tumbling i.e. results in the freezing out of the Brownian relaxation. 

\begin{figure}[h!]
    \centering
    \begin{subfigure}{\linewidth}
        \centering
        \includegraphics[width=\linewidth]{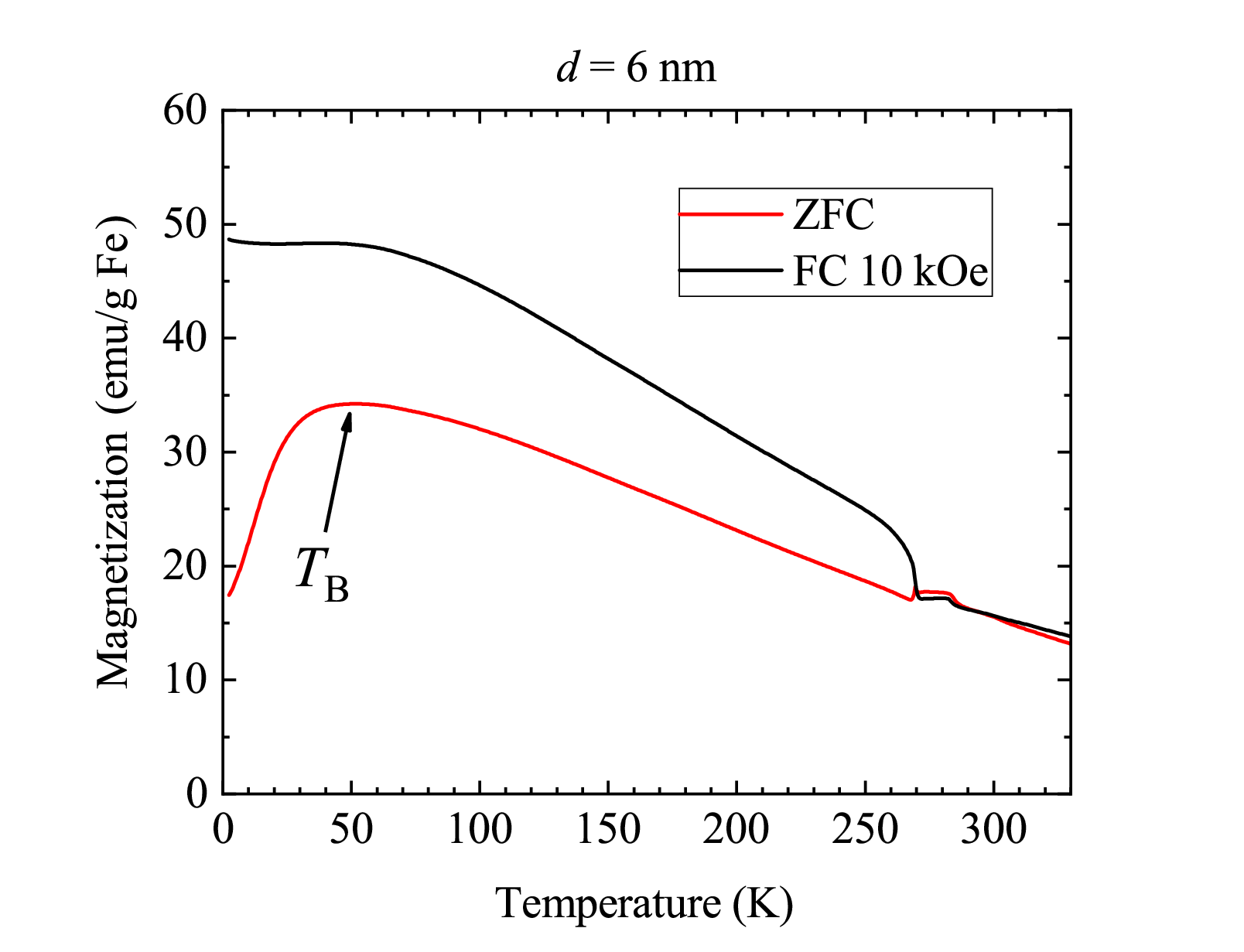} 
   \end{subfigure}
   \hfill
    \begin{subfigure}{\linewidth}
        \centering
        \includegraphics[width=\linewidth]{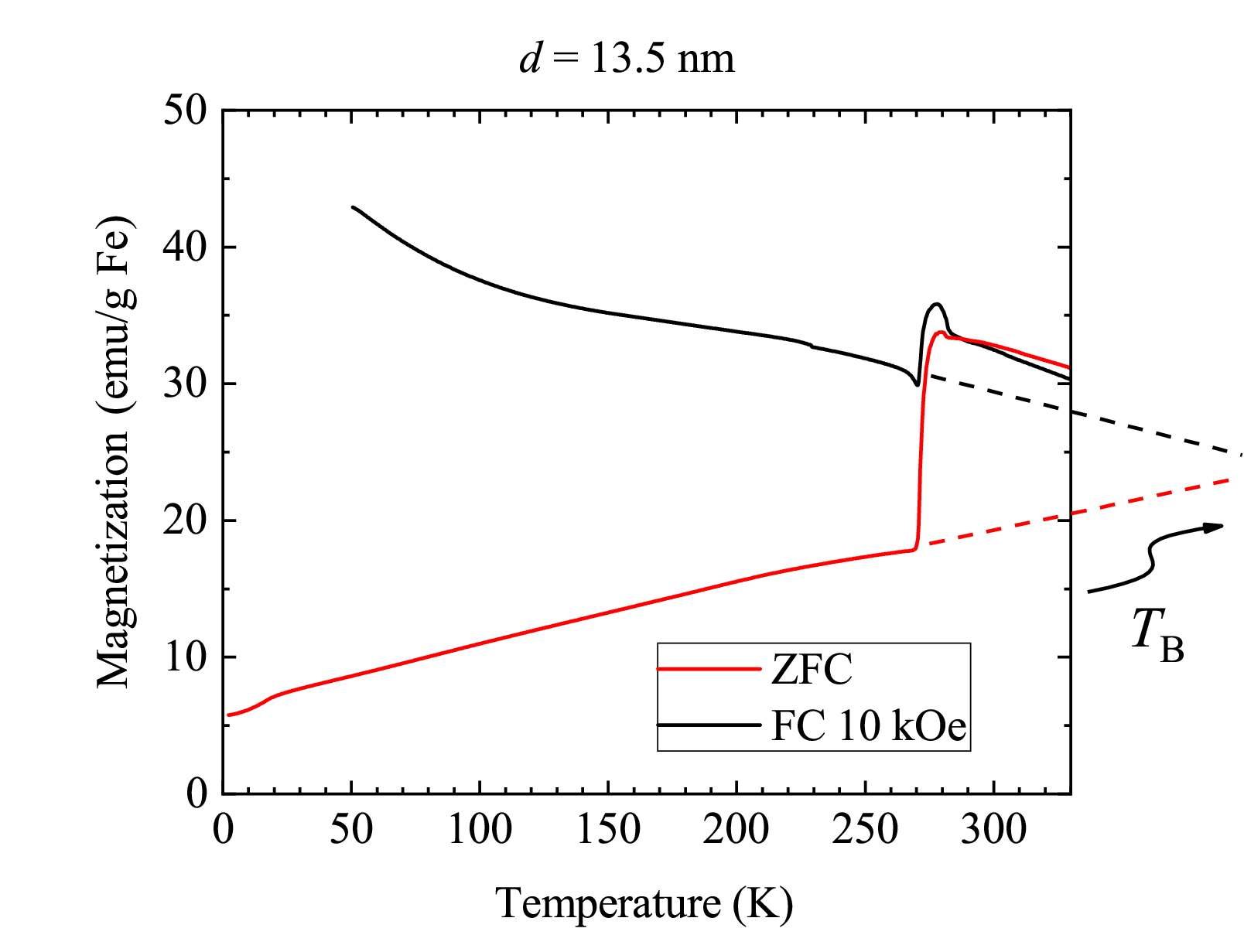}
        \end{subfigure}
    \caption{Zero field-cooled and field-cooled (in 10 kOe) magnetization measurements for the 6-nm and 13.5-nm samples. Both measurements were performed upon warming at 100 Oe. The approximate position of $T_{\text{B}}$ is indicated for the 6-nm sample, however, it lies above 300 K for the 13.5-nm sample. For the latter sample, an extrapolation to the low-temperature ZFC data is shown with a dashed line, and a higher-than-room temperature $T_{\text{B}}$ is indicated.}
    \label{zfc_fc}
\end{figure}

The above observations and the drastically different behavior for the two samples are further supported by a series of field-cooled (FC) and zero-field-cooled (ZFC) experiments, which are shown in Figure \ref{zfc_fc}. In both cases, the measurements were performed upon heating at 100 Oe after cooling either in zero field or at 10 kOe. The curves are offset manually (in both cases by not more than 5 emu/g Fe) so that the data match at 300 K. This is necessary as the amount of trapped flux is unknown following the field cooling, as mentioned in the Experimental section. We cannot exclude either that in the liquid form, the ferrofluid moves in the sealed capillary, which also affects the measurements.

In the ZFC experiment, the two samples show markedly different behavior: there is a large jump in the magnetization (about 50 \%) for the 13.5-nm sample around 273 K but the magnetization in the 6-nm sample shows a negligible change at this temperature. This observation is fully compatible with the expected \emph{dominance} of the Brownian relaxation for the 13.5-nm sample \cite{garaio}. The magnitude of the jump is comparable in the FC experiment for both samples, an effect whose explanation is less evident: one would expect that the 6-nm sample is not affected at all by the water freezing. The observation may be due to a differing preferred direction for the small nanoparticles in 10 kOe.

For the 6-nm sample, a maximum in the ZFC data magnetization is observed around 50 K, which is usually associated with the so-called blocking temperature \cite{blocking_th}, $T_{\text{B}}$. This corresponds to the blocking of the rotation of sample magnetization, i.e. below this point, the Néel process is also hindered and the sample no longer shows a superparamagnetic behavior. For the 13.5-nm sample the blocking temperature is expected to be above room temperature \cite{blocking_size}. In fact, extrapolation of the low-temperature ZFC for the 13.5-nm sample (dashed line in Figure \ref{zfc_fc}) shows a merger to the FC data i.e. $T_{\text{B}}$ above room temperature. Clearly, the blocking temperature cannot be observed in the 13.5-nm sample above 300 K as the nanoparticle tumbling prevails. However, this blocking temperature would be observed for the 13.5-nm sample if the nanoparticles were embedded in a solid matrix.


\begin{figure}[h!]
    \begin{subfigure}{\linewidth}
        \centering
        \includegraphics[width=\linewidth]{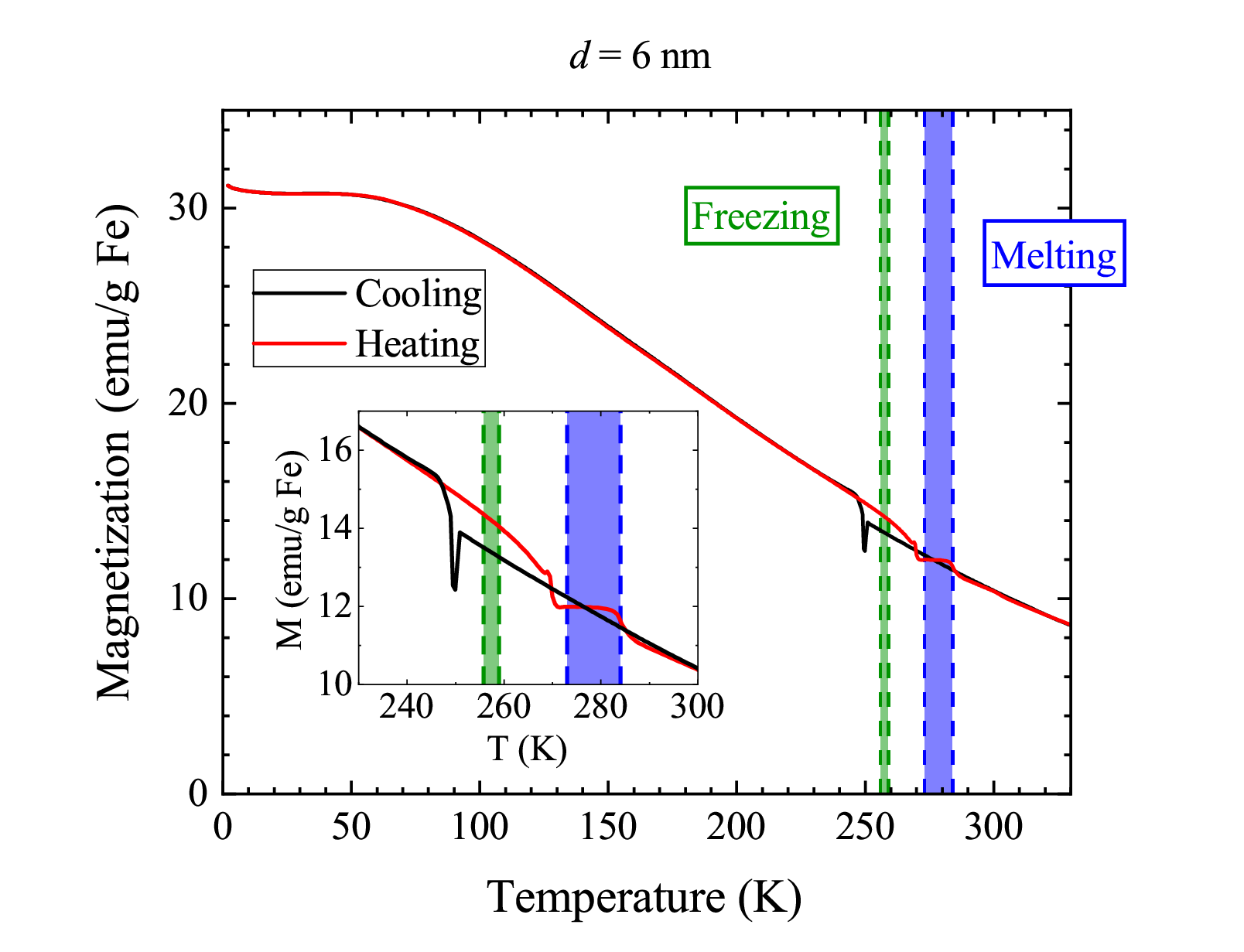} 
   \end{subfigure}
   \hfill
    \begin{subfigure}{\linewidth}
        \centering
        \includegraphics[width=\linewidth]{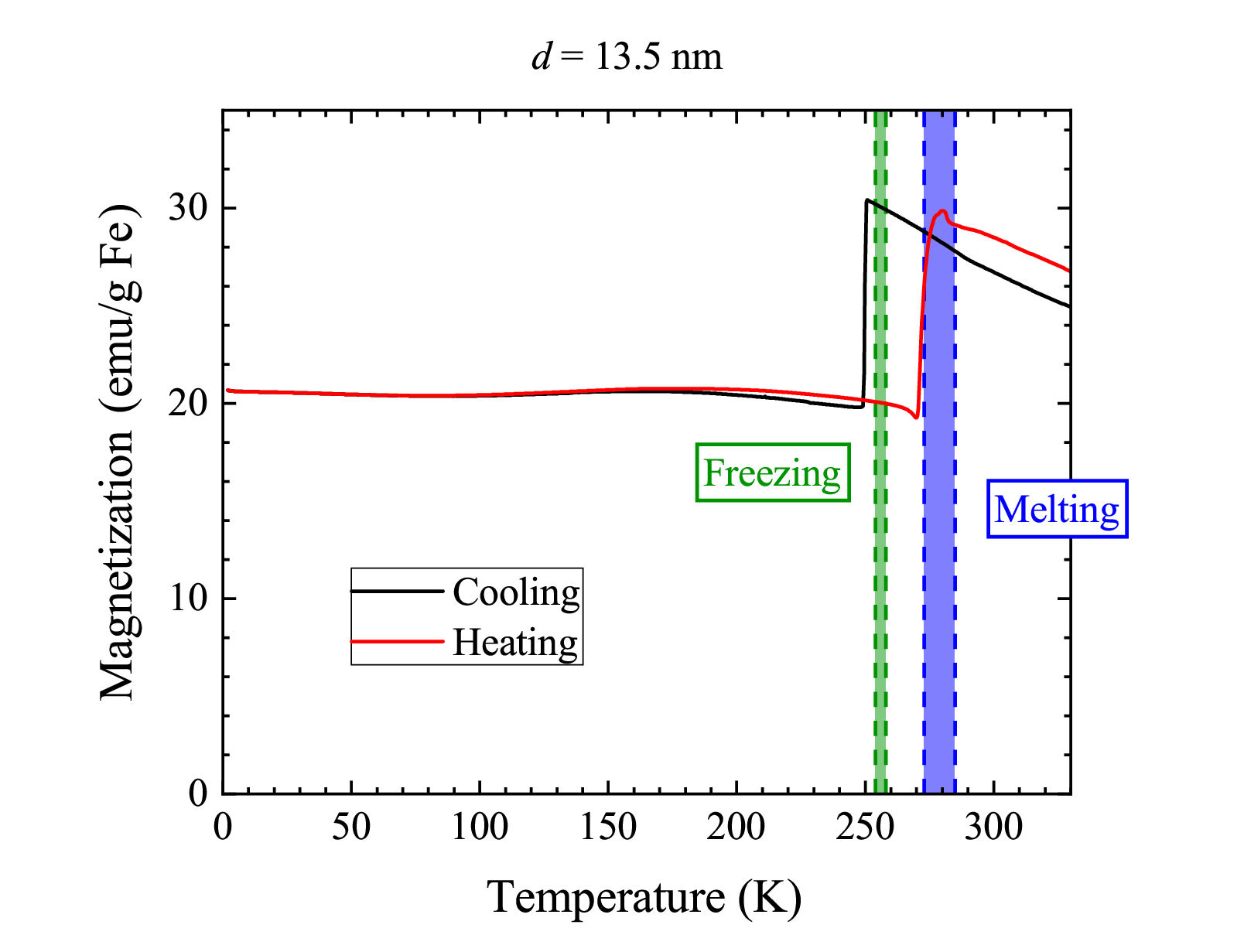}
        \end{subfigure}
    \caption{Thermal hysteresis studied by monitoring the magnetization in a moderate magnetic field (100 Oe) as a function of temperature in cooling and warming for the 6-nm sample (top) and the 13.5-nm sample (bottom). The temperature \emph{ranges}, where freezing and melting occur, are obtained from the DSC studies and are indicated by shaded areas. The inset of the }
    \label{100Oe}
\end{figure}

In addition to the conventional FC and ZFC measurements, we found that thermal hysteresis measurements in moderate magnetic fields (100 Oe) also provide valuable information into the magnetization behavior. In contrast to the high magnetic field in the FC studies (10 kOe), the applied 100 Oe is insufficient to fully orient the nanoparticle magnetization and it is rather a "readout" magnetic field. Therefore it allows us to read out the magnetic state of the nanoparticles without strongly affecting it. The data is shown in Figure \ref{100Oe}: in the measurement, we first cooled and then warmed the sample in the same unchanged 100 Oe magnetic field.

The figure shows shaded areas for the freezing and melting of the same sample suspensions using differential scanning calorimetry, DSC. The details of the DSC analysis are provided in the Supporting Information. A freezing-melting hysteresis is expected given the first-order nature of this phase transition. The DSC method delivers range, rather than exact temperatures for the onset of these processes. As a cautionary note, we mention that melting/freezing in the vicinity of the nanoparticles may not necessarily occur at the same temperature as melting/freezing in the bulk of the solvent \cite{alapkut_puerto,jackpot}, however in our measurement, the two effects appear approximately at the same temperature.

We observe a thermal hysteresis for both types of samples between the freezing and melting temperatures. However, it is much larger for the 13.5-nm sample as expected when the Brownian relaxation dominates the sample magnetization. The magnetization in the 13.5-nm sample is higher in the liquid state which means that the Brownian relaxation allows for the rotation of the nanoparticle along a preferred direction of the nanoparticle magnetization with respect to the external field. 

The data also shows an interesting, albeit yet unexplained behavior for the 13.5-nm sample. One expects the magnetization to increase as the solvent melts but that it stays constant when the solvent freezes since the particles could rotate to better align their magnetic moments with the applied field in the liquid. In contrast to this expectation, the magnetization decreases suddenly below the freezing of water. We surmise that the Brownian process allows for increased magnetization due to the dynamical alignment of the particles.

We also studied the concentration dependence of this effect (data shown in the Supporting Information) to find out whether the particle-particle interactions may give rise to this but we observed none. We discuss below that the vanishing of a part of the sample magnetization upon freezing of the solvent for the 13.5-nm sample corroborates the AC susceptibility measurements. Our observations also indicate that the low-temperature ground state of the material, when cooled at 100 Oe is insensitive to the thermal history.

An excellent benchmark method to reveal the Brownian relaxation is the type of data shown in Figure \ref{100Oe}: it only requires the application of a moderate magnetic field, whose magnitude does not need to be precisely calibrated. 

\begin{figure}[h!]
    \centering
    \begin{subfigure}{\linewidth}
        \centering
        \includegraphics[width=\linewidth]{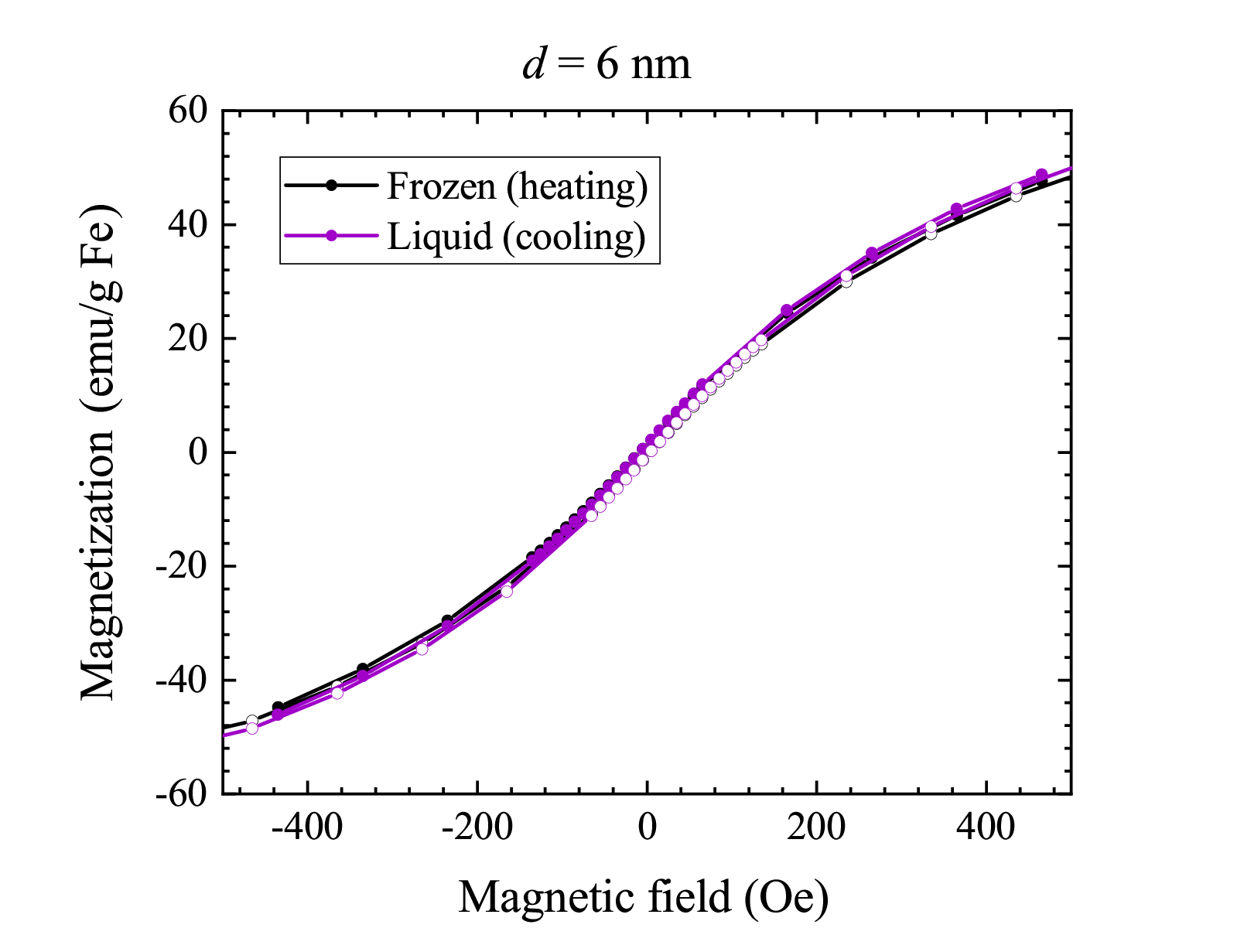} 
   \end{subfigure}
   \hfill
    \begin{subfigure}{\linewidth}
        \centering
        \includegraphics[width=\linewidth]{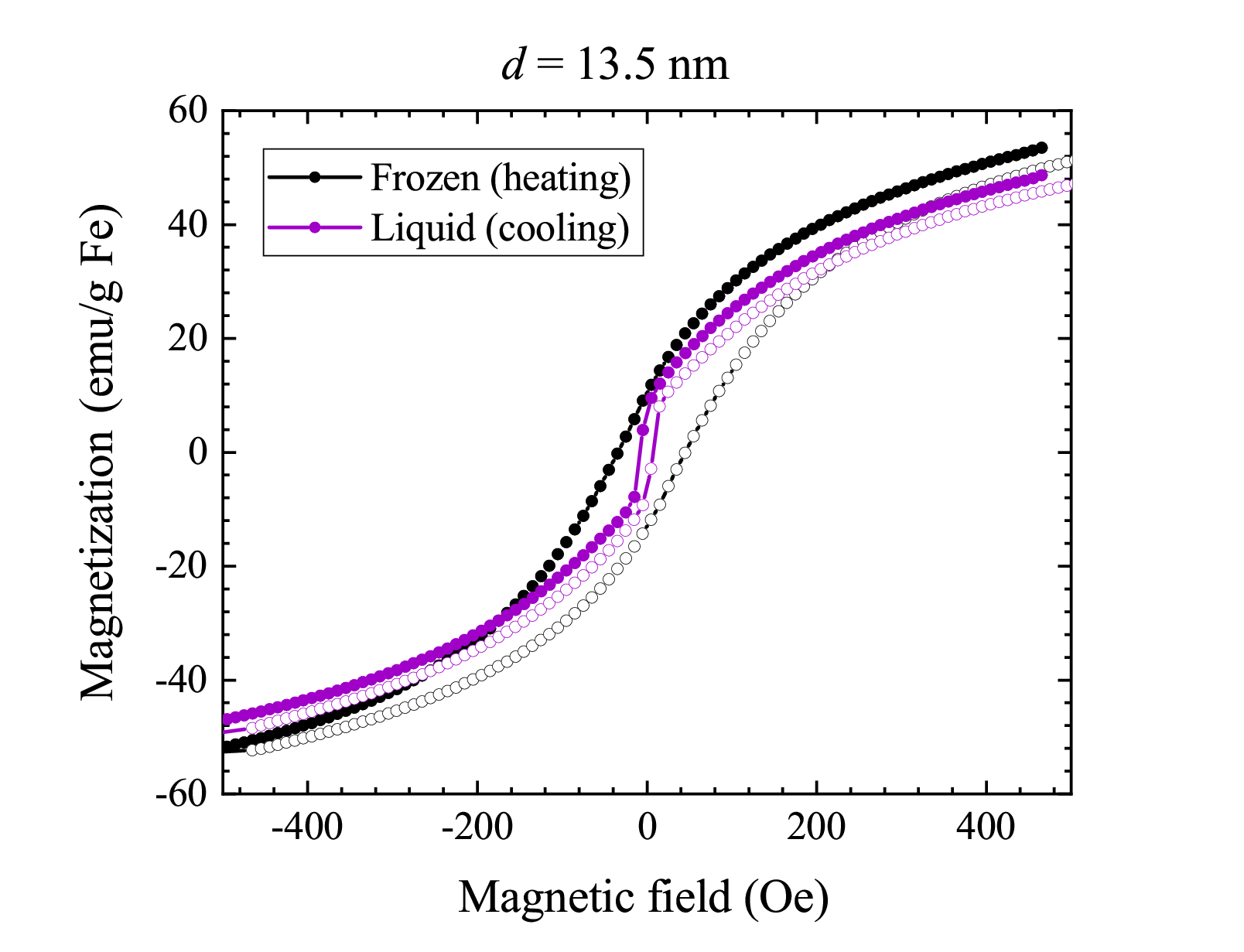}
        \end{subfigure}
    \caption{Magnetic hysteresis curves at 270 K in liquid and solid states in the case of the 6-nm sample (top) and the 13.5-nm sample (bottom). Liquid state: purple curves, frozen state: black curves. Data denoted by full symbols were recorded while decreasing the magnetic field, whereas open symbols stand for data measured at increasing magnetic field.}
    \label{phase}
\end{figure}

The thermal hysteresis studies indicate that there exists a temperature range (between about 250-280 K) where both the liquid and solid state can be studied at the \emph{same} temperature depending on the thermal history. To this end, we studied magnetic hysteresis at 270 K where both phases are realized upon cooling or warming, and the result is shown in Figure \ref{phase}. No magnetic hysteresis is observed for the 6-nm sample in either case, therefore the magnetization direction is unaffected by the state of the surroundings for small-sized nanoparticles. This is significantly different for the 13.5-nm sample where a pronounced hysteresis appears when the solvent is frozen. This is in full agreement with the dominance of the Brownian process for this particle size.

\begin{figure}[h!]
    \centering
    \begin{subfigure}{\linewidth}
        \centering
        \includegraphics[width=\linewidth]{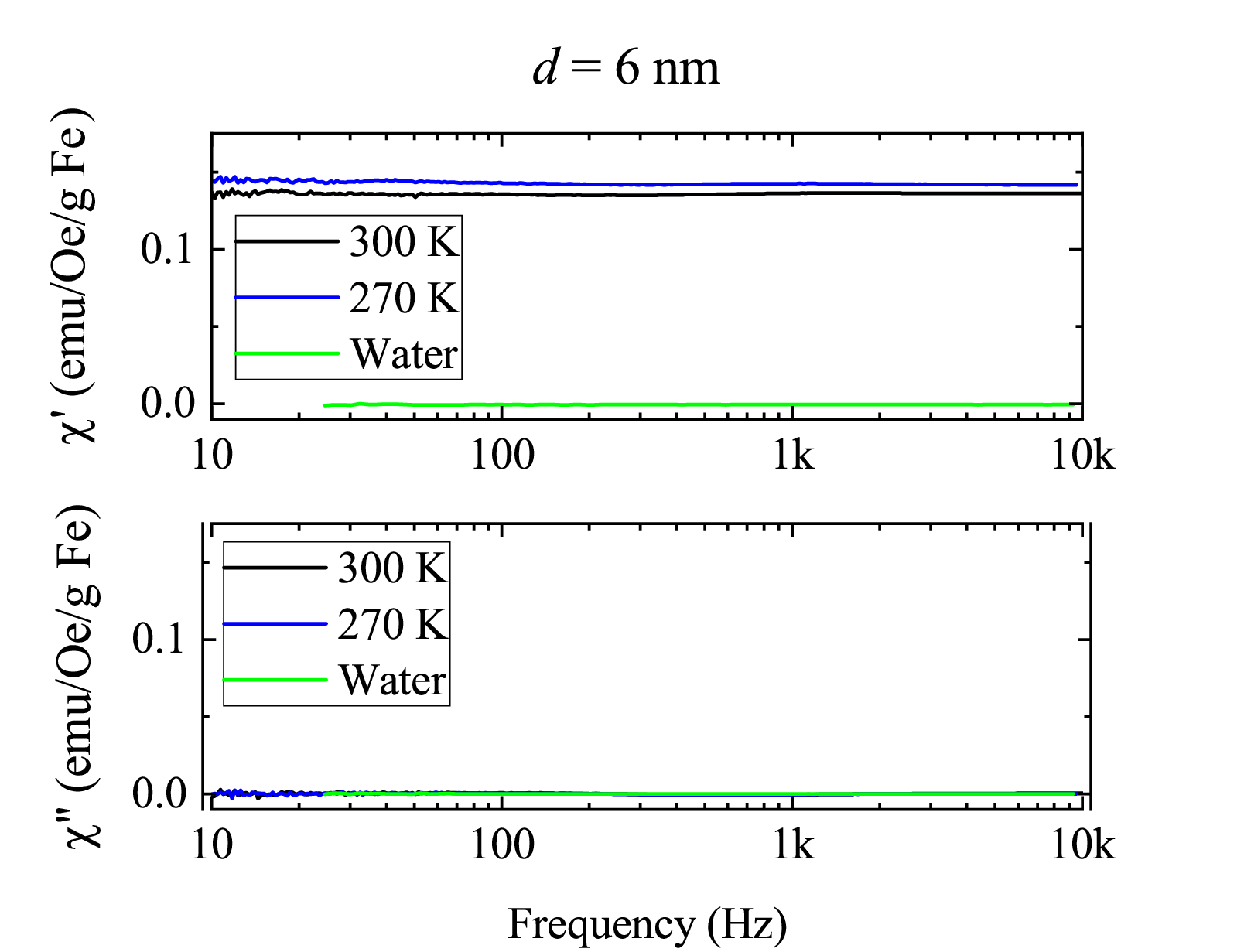} 
   \end{subfigure}
   \hfill
    \begin{subfigure}{\linewidth}
        \centering
        \includegraphics[width=\linewidth]{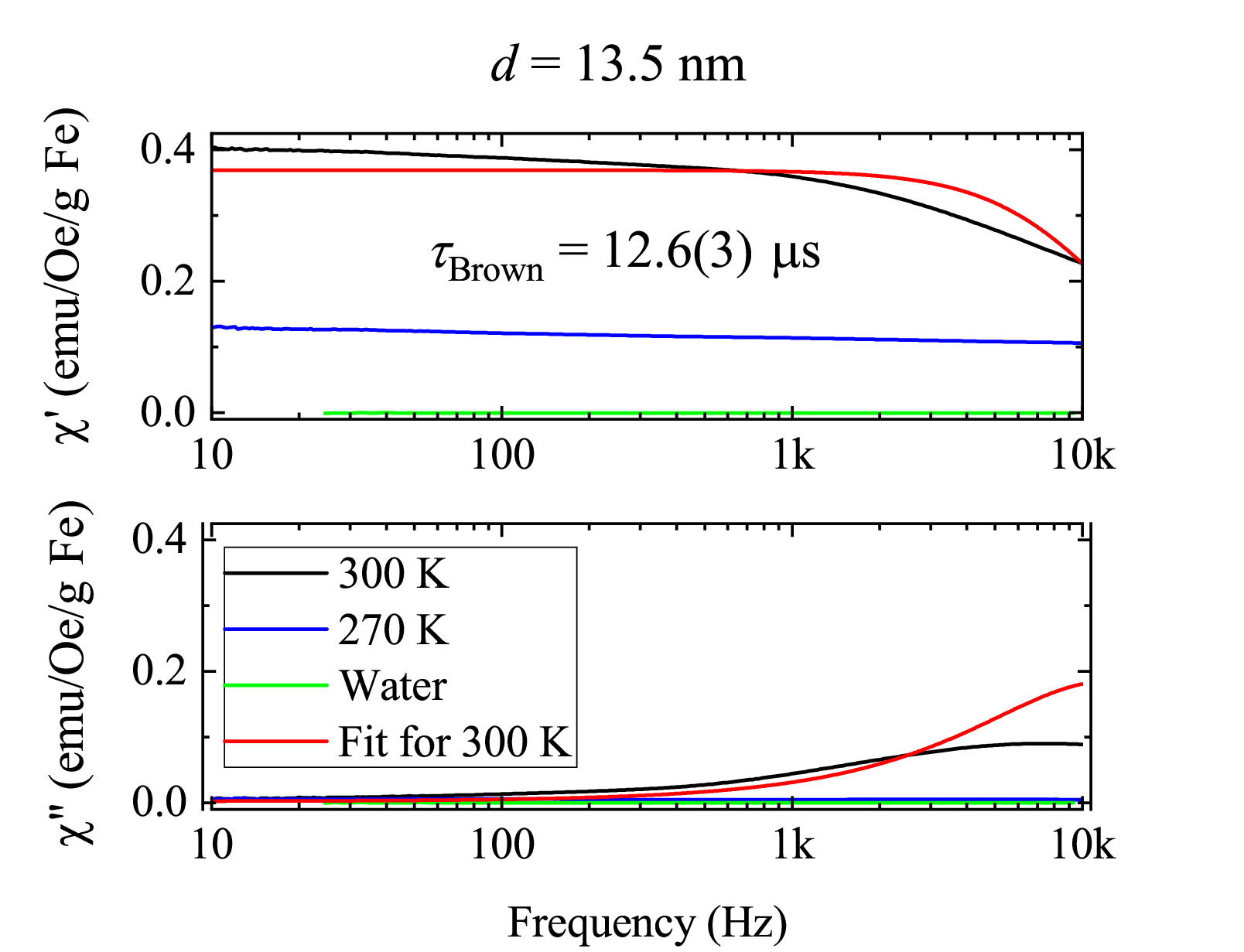} 
        \end{subfigure}
    \caption{Frequency spectra of the real and imaginary parts of AC magnetic susceptibility for a few temperatures across the water freezing/melting for both samples. For the 13.5-nm sample, a Lorentzian fit is provided from a mono-exponential Debye model.}
    \label{acms}
\end{figure}

 Figure \ref{acms} shows the frequency-dependent real and imaginary parts of the AC magnetic susceptibility at various temperatures for both samples. The complex AC magnetic susceptibility spectra are expected to follow a Lorentzian curve
 \cite{Review_Garaio,Review_ortega_pankhurst,Review_Wu} from a mono-exponential Debye model as:

\begin{equation}
\widetilde{\chi}\left(\omega\right)=\chi'\left(\omega\right)+\mathrm{i}\chi''\left(\omega\right)=\\
\chi_0\left[\frac{1}{1+\omega^2\tau^2}+\mathrm{i}\frac{\omega\tau}{1+\omega^2\tau^2} \right]
\end{equation}
herein, $\chi_0$ is the static susceptibility, the angular frequency is $\omega = 2\pi f$, and the relaxation time, $\tau$ is a combination of the Néel ($\tau_{\text{Néel}}$) and Brownian ($\tau_{\text{Brown}}$) relaxation times: 
\begin{equation}
\frac{1}{\tau}=\frac{1}{\tau_{\text{Néel}}}+\frac{1}{\tau_{\text{Brown}}}
\end{equation}

The linear response theorem dictates that the real and imaginary parts are Hilbert transforms of one another.

For the smaller diameter nanoparticles (6-nm sample) freezing of the solvent (300 K vs. 270 K), does not change the magnetic susceptibility appreciably: $\chi '$ slightly increases, which is consistent with the steepening of the hysteresis curves in Figure \ref{hist} in the small magnetic field regime. There is no observable sign of relaxation in this frequency range, the characteristic frequency of Néel relaxation processes (i.e. $1/\tau_{\text{Néel}}$) is typically much larger \cite{Review_Garaio,Review_ortega_pankhurst,Review_Wu}. 

For the larger nanoparticles (13.5-nm sample) in the liquid phase, there is a visible Debye function corresponding to Brownian relaxation with a relaxation time $\tau_{\text{Brown}}$. Upon freezing, this contribution vanishes and the curves become similar in shape and magnitude to the ones recorded for the 6-nm sample. This implies the presence of a residual Néel process even when the sample is frozen.

A global Lorentzian fit for data taken at 300 K for the 13.5-nm sample was performed and the result is shown in Figure \ref{acms} along with the fit parameters. The fit is not accurate, probably due to a distribution in the particle sizes (see Supporting Information). Clearly, in our frequency range, we cannot determine the frequency of the Néel process. Our fit gave a relaxation time of $\tau_{\text{Brown}}=12.6\,\mathrm{\mu s}$. The seminal paper of Ortega and Pankhurst \cite{Review_ortega_pankhurst} tabulates that such a Brownian relaxation time corresponds to a particle size of about 12 nm based on relaxation time calculations performed for similar nanoparticle systems to the ones investigated here. This result is in good agreement with our nominal particle size of 13.5 nm. We also note that the Brownian relaxation is determined by the hydrodynamic diameter, which may be different from the apparent nanoparticle diameter (see Supporting Information).

The vanishing of the Brownian relaxation and its sensitivity to the temperature has a direct consequence for the application of nanoparticles in e.g. nanomagnetic hyperthermia. Therein, the knowledge of the frequency-dependent magnetic susceptibility is crucial to determining the optimal irradiation conditions. In turn, this information may provide a spectroscopic fingerprint on the nature of the surrounding tissue: when the nanoparticles are immobile, the AC magnetization frequency spectrum is significantly altered. 

\section{Conclusions}

We observed drastic diameter-dependent differences in the magnetometric and calorimetric response of well-characterized ferrofluidic magnetite nanoparticle water suspensions. Namely, we identified the fingerprints of both magnetic and thermal hysteresis on the nanoparticle magnetism when the Néel or the Brownian relaxation processes dominate. These observations may serve as benchmarks for sample characterization solely from the magnetization data. We identified that a low magnetic field measurement through the solvent freezing/melting temperature provides the simplest means to characterize whether the Brownian relaxation dominates the magnetic properties. We also identified the fingerprints of the different relaxation processes in the frequency-dependent magnetization data, which has direct therapeutic relevance for medical applications.

\section{Supporting Information}

A detailed description of the sample preparation method is given, then the results of the sample size characterization are summarized. This is followed by the summary of the calculation of converting magnetization units between SI and CGS as well as between the values normalized to the mass of iron and magnetite. There is a detailed calculation of approximate sample sizes based on room temperature magnetic hysteresis data, whose results can be found in the main text. The concentration dependence of the observed effects is investigated with an emphasis on the comparison of zero-field-cooled and field-cooled magnetization. The particle size dependence is also looked at similarly. Finally, the results of Differential Scanning Calorimetry and the determination of the freezing and melting temperature ranges are also detailed.

\section*{Acknowledgements}
The Authors acknowledge the kind help of Eva Céspedes and José Luis Martínez with the experiments. This work was supported by the Hungarian National Research, Development and Innovation Office (NKFIH) Grants 2022-2.1.1-NL-2022-00004, K137852, FK138501, TKP 2021-NVA-04, and the V4-Japan Joint Research Program (BGapEng, 2019-2.1.7-ERA-NET-2021-00028), and TED2021-129254B-C21 of the Spanish MCIN. 

\onecolumngrid

\cleardoublepage

\makeatletter

\renewcommand{\thefigure}{S\@arabic\c@figure}
\renewcommand{\thetable}{S\arabic{table}}

\makeatother

\setcounter{figure}{0}

\appendix

\sloppy

\section{Supporting Information}

\subsection{Sample preparation and characterization}

The nanoparticles were obtained following the coprecipitation protocol described by Massart \cite{Massart_np_preparation}, by introducing small modifications to control particle size between 6 and 14 nm.
The synthesis of Fe$_3$O$_4$ magnetite nanoparticles was carried out by mixing 488 mL of a ferrous-ferric solution, FeCl$_3\cdot$6H$_2$O (0.09 mol) (27\% purity, $d=$ 1.26 kg/L, MW $=$ 162.21 g/mol, Sigma-Aldrich) and FeCl$_2\cdot$4H$_2$O (0.045 mol) ($\geq$99\% purity, MW $=$ 198.81 g/mol, Sigma-Aldrich), with 75 mL of NH$_4$OH (28.0-30.0\% purity, $d=$ 0.9 g/mL, MW $=$ 35.05 g/mol, Sigma-Aldrich). 
The precursors were added and magnetically stirred. Subsequently, it was allowed to cool to room temperature and washed three times by magnetic decantation with distilled water. To obtain larger nanoparticles, the sample was subjected to a heat treatment at 90 $^\circ C$ for 1 hour before washing.

The precipitate was subjected to an acid treatment to improve the colloidal stability of the nanoparticles.
To do this, 300 mL of 2M HNO$_3$ nitric acid (65\% purity, $d=$ 1.68 g/cm, MW $=$ 404.00 g/mol, Sigma-Aldrich) were added to the nanoparticles and kept stirring for 15 min.
After that time, the supernatant was discarded by magnetic decantation and 75 mL of 1M Fe(NO$_3$)$_3$ iron nitrate ($\geq$99\% purity, $d=$ 1.37-1.41 g/mL, MW $=$ 63) and 130 mL of distilled water were added.
The sample was boiled for 30 min while stirring, then it was allowed to cool to room temperature and treated again with 300 mL of 2 M HNO$_3$ for 15 min.
The resulting nanoparticles were washed 3 times by magnetic decantation with acetone.
Subsequently, the nanoparticles were resuspended in 90 mL of distilled water.
The acetone was completely removed with the help of the Rotavapor at 60 $^\circ $C and 556 mbar (Rotation evaporator Laborota 4011, Heidolph).

\begin{figure}[h!]
    \centering
    \includegraphics[width=0.4\linewidth]{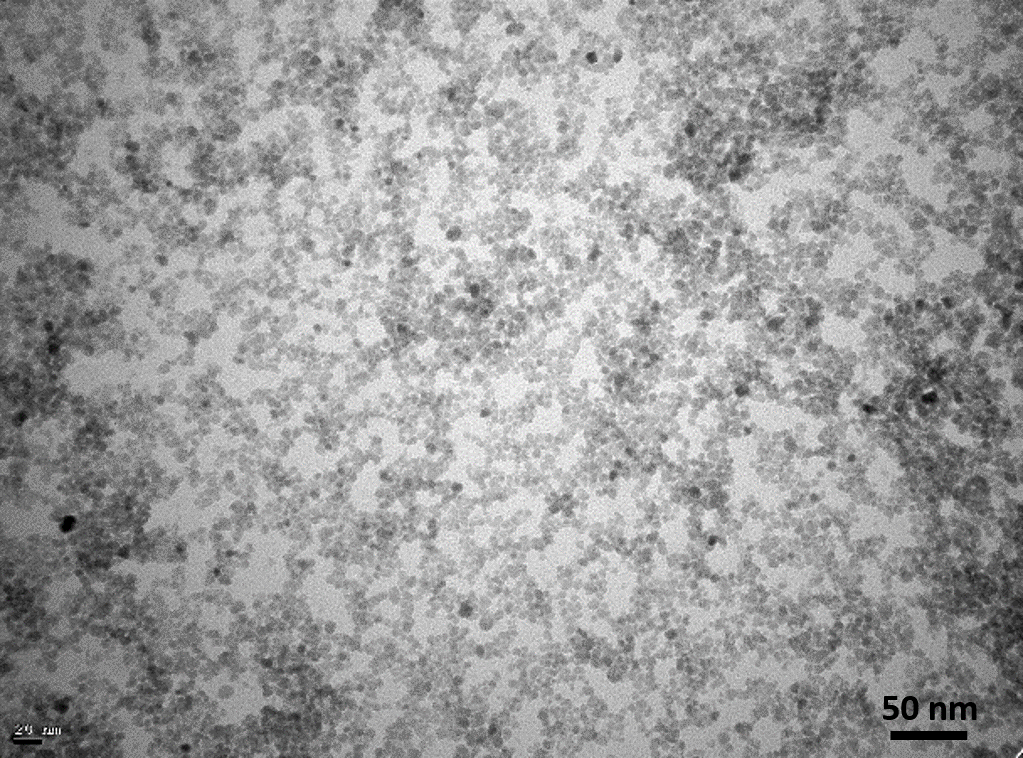}
    \caption{TEM image of the 6-nm sample.}
    \label{SM_6nm_TEM}
\end{figure}

\begin{figure}[h!]
    \centering
    \includegraphics[width=0.4\linewidth]{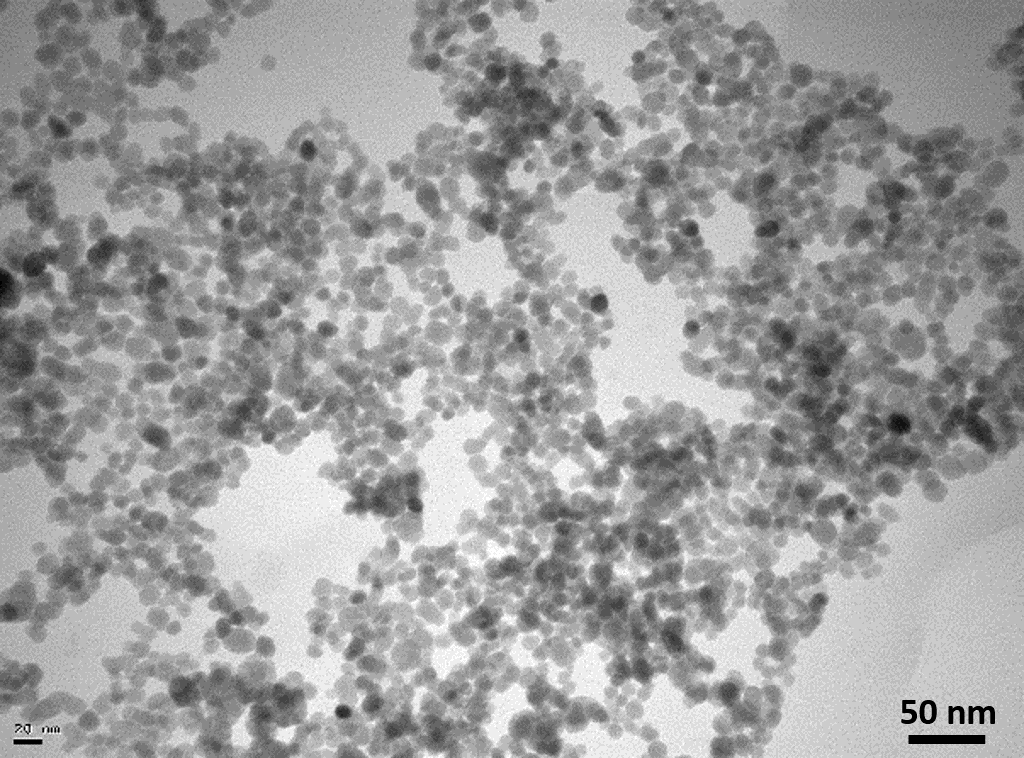}
    \caption{TEM image of the 13.5-nm sample.}
    \label{SM_13nm_TEM}
\end{figure}

\begin{figure}[h!]
    \centering
    \includegraphics[width = 0.5\linewidth]{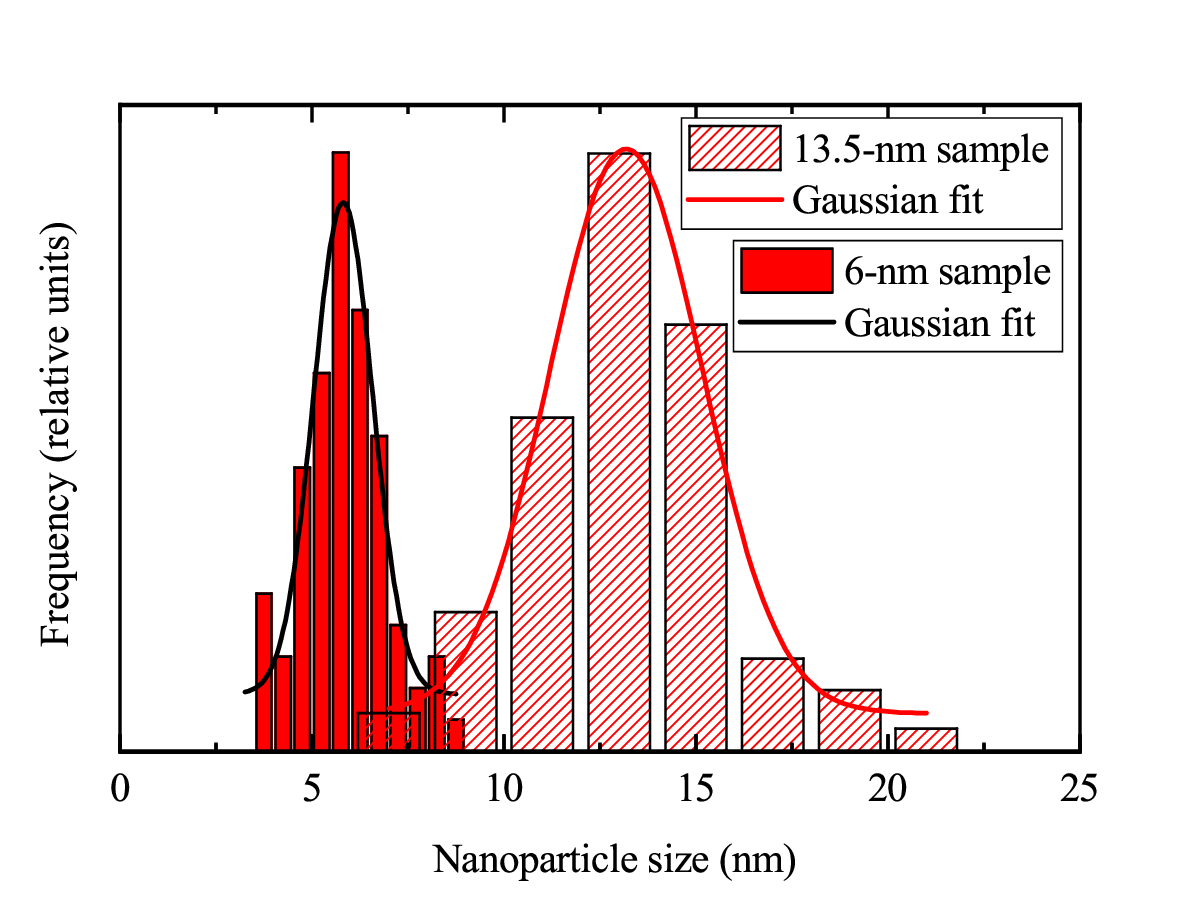}
    \caption{Histogram of particle sizes based on the TEM images of the 6-nm and 13.5-nm samples along with Gaussian fits.}
    \label{SM_sample_prep_histogram}
\end{figure}

Particle sizes were measured with TEM, and the recorded images are depicted in Figure \ref{SM_6nm_TEM} (6-nm sample) and \ref{SM_13nm_TEM} (13.5-nm sample). Based on these images, particle size histograms were created (Figure \ref{SM_sample_prep_histogram}), and Gaussian fits were performed. As a result, particle size data were obtained as 5.8$\pm$1.5 nm for the 6-nm sample, and 13.2$\pm$4 nm for the 13.5-nm sample.

Hydrodynamic sizes in acid (pH = 3) were measured with dynamic light scattering, resulting in 40 nm (PDI: 0.13) for the 6-nm sample and 76 nm (PDI: 0.17) for the 13.5-nm sample. As expected, the hydrodynamic sizes are much larger than the nominal ones, affecting the dynamics of Brownian relaxation.

\subsection{Magnetization normalization and conversion}

The mass magnetization in CGS units (emu/g) can be conveniently converted into the SI (volume) magnetization units of A/m using that 1 emu $=10^{-3}$ Am$^2$ (NIST link: https://www.nist.gov/document/magneticunitspdf). Correspondingly, for the volume magnetization, we obtain 1 emu/cm$^3 = 10^3$ A/m. As a result, for magnetite with 80 emu/g Fe$_3$O$_4$ and density 5.170 g/cm$^3$, we obtain 413 kA/m \cite{Magnetite_80emu_per_g}.

When normalized by the Fe mass, the emu/g Fe values have to be multiplied by 0.723 to obtain emu/g Fe$_3$O$_4$ values.

\subsection{Magnetic particle size evaluation based on the Langevin function slope}

\begin{figure}[h!]
    \centering
    \begin{subfigure}{0.49\linewidth}
        \centering
        \includegraphics[width=\linewidth]{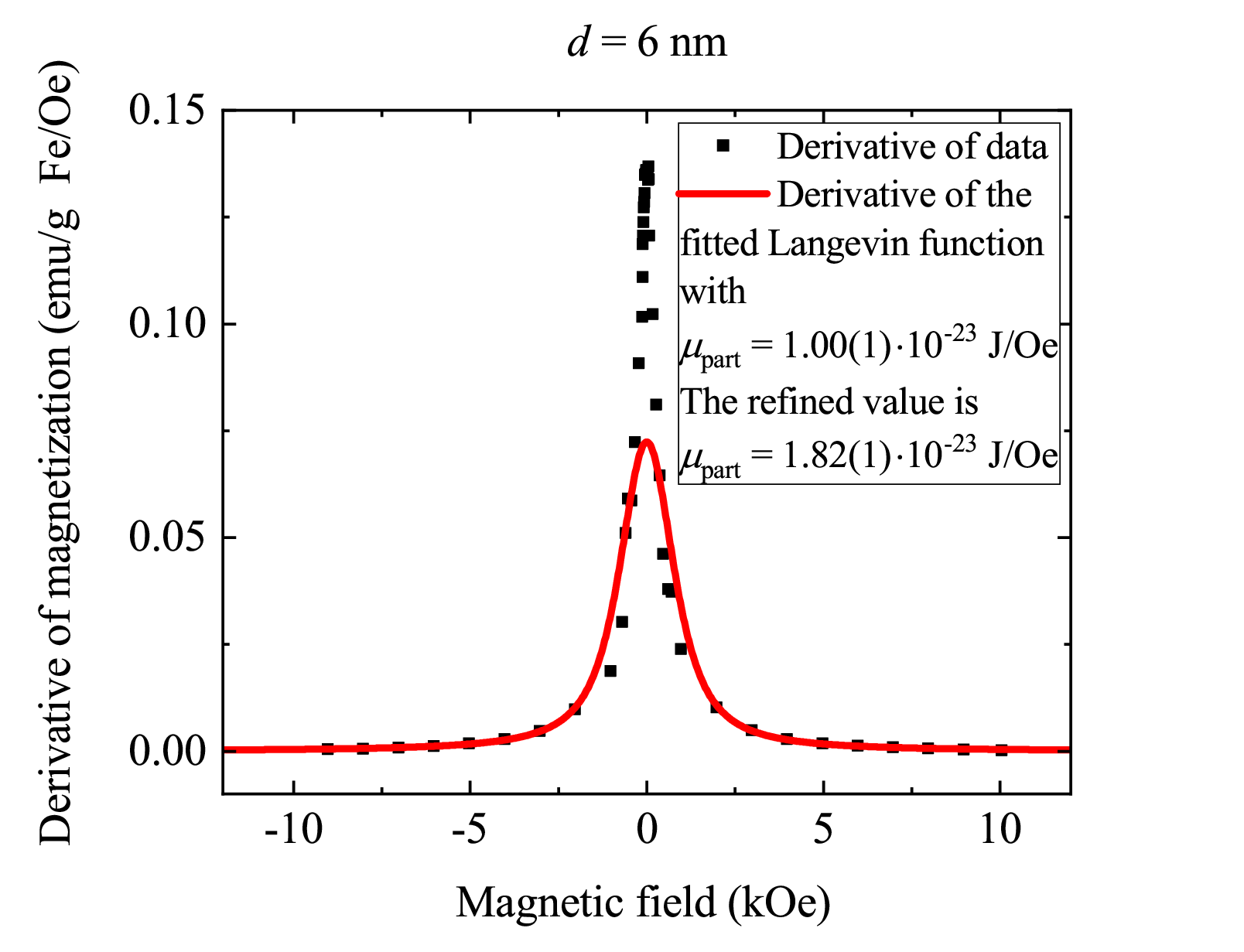} 
   \end{subfigure}
   \hfill
    \begin{subfigure}{0.49\linewidth}
        \centering
        \includegraphics[width=\linewidth]{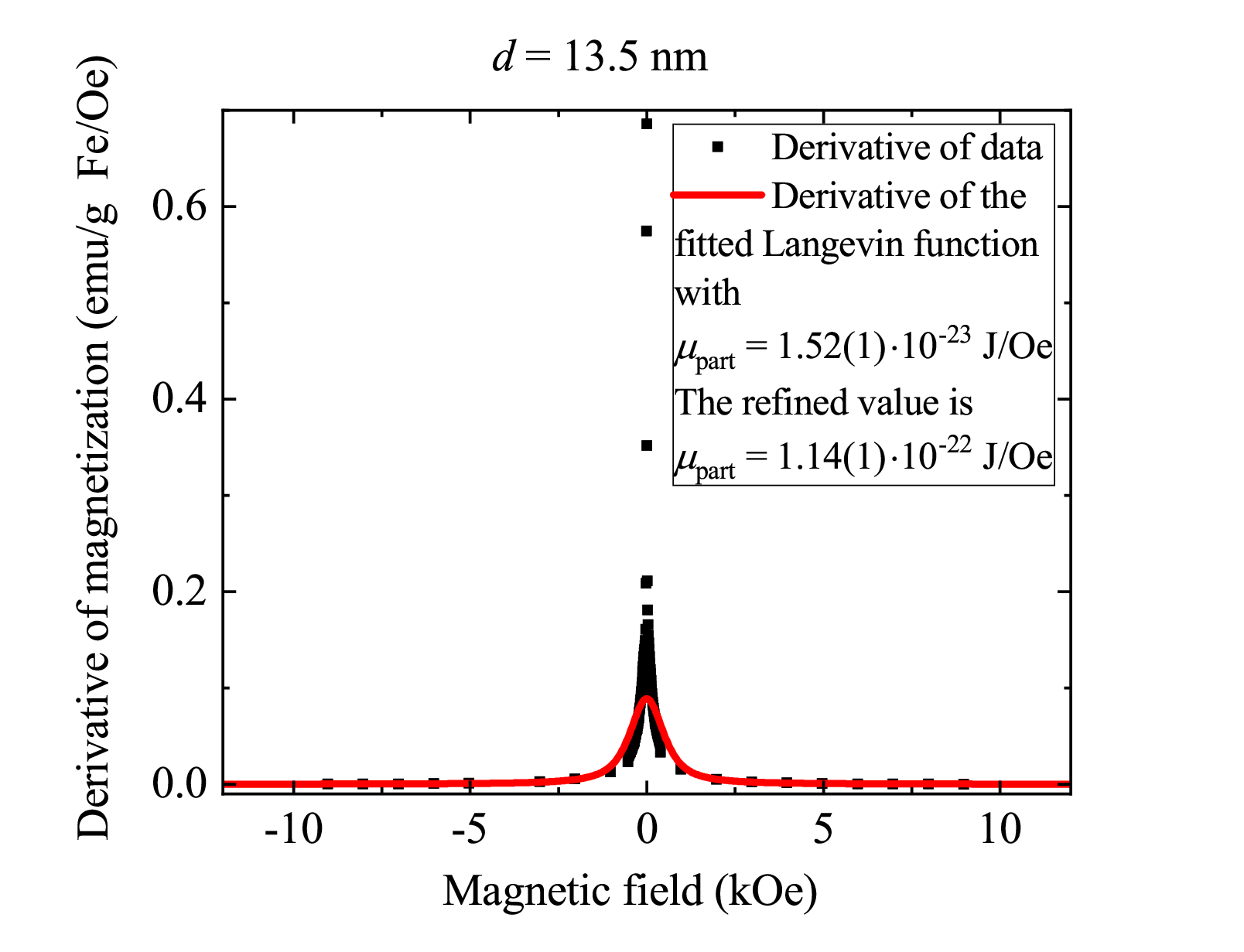}
        \end{subfigure}
    \caption{Derivative of the magnetization curves for the two types of magnetite samples recorded at 300 K: containing 6-nm particles (top) and 13.5-nm particles (bottom). Solid red curves show the derivative of the fitted Langevin curves. Clearly, the derivative of the fits is below that of the data. This indicates that the $\mu_\text{part}$ values obtained from the Langevin fits are below that obtained from matching the $B=0$ slope. The latter gives rise to significantly increased values of $\mu_\text{part}$ as discussed in the main text.}
    \label{langevin_derivative}
\end{figure}

The main text showed fits to the room temperature magnetization data using the Langevin function. This function has two parameters: the saturated magnetization, $M_\text{s}$, and the particle magnetic moment, $\mu_\text{part}$. The latter is essentially a horizontal scaling factor, which is in turn sensitive to the particle size. The nature of the fits is such that it is more sensitive to the value of $M_\text{s}$ than to $\mu_\text{part}$. However, $\mu_\text{part}$ can be also determined from a numerical derivative of the data as shown in Figure \ref{langevin_derivative}.

The figure shows a comparison between the derivative of the original data and that of the fitted Langevin functions. Clearly, the derivative of the Langevin functions is below for both types of samples. In order to match the two types of data, $\mu_\text{part}$ has to be increased by about a factor 2 for the 6 nm sample and by a factor of about 6 for the 13.5 nm sample. As the main text discusses, these increased $\mu_\text{part}$ values give rise to a more realistic magnetic particle size.  

\newpage
\subsection{Concentration dependence of the observed effects}

The main text discusses results on samples with 3 mg/mL concentrations. In the following, we discuss the effect of the varying concentrations. The major result can be summarized as follows: the observed behaviors do not show any concentration dependence for 10 and 20 mg/mL but changes start to appear for 30 mg/mL. 

\begin{figure}[h!]
    \includegraphics[width=0.5\linewidth]{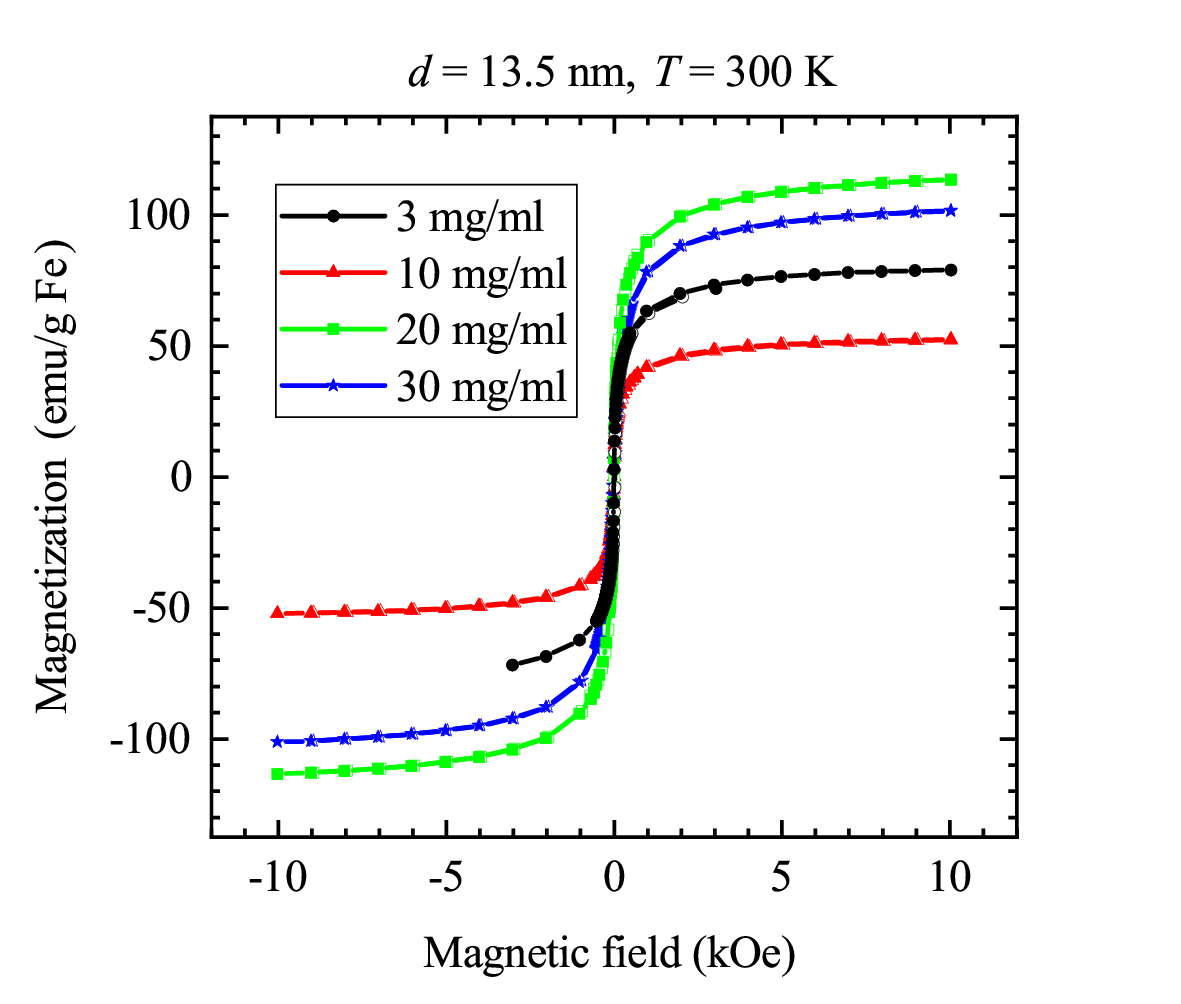}
    \caption{Magnetic field-dependent magnetization curves for the 13.5 nm sample for 4 different concentrations. The different saturated values are probably due to the errors in the estimate of the Fe concentration.}
    \label{SM_13x5_saturation}
\end{figure}

The error of the Fe amount estimate is significant (which can be traced back to the large error of the mass measurement): the smallest value of $M_\text{sat}$ is 49.32 emu/g Fe for the 10 mg/mL sample, while it is 110 emu/g Fe for the 20 mg/mL sample with no apparent systematic dependence of these values on the concentration.

\begin{figure}[h!]
    \includegraphics[width=0.5\linewidth]{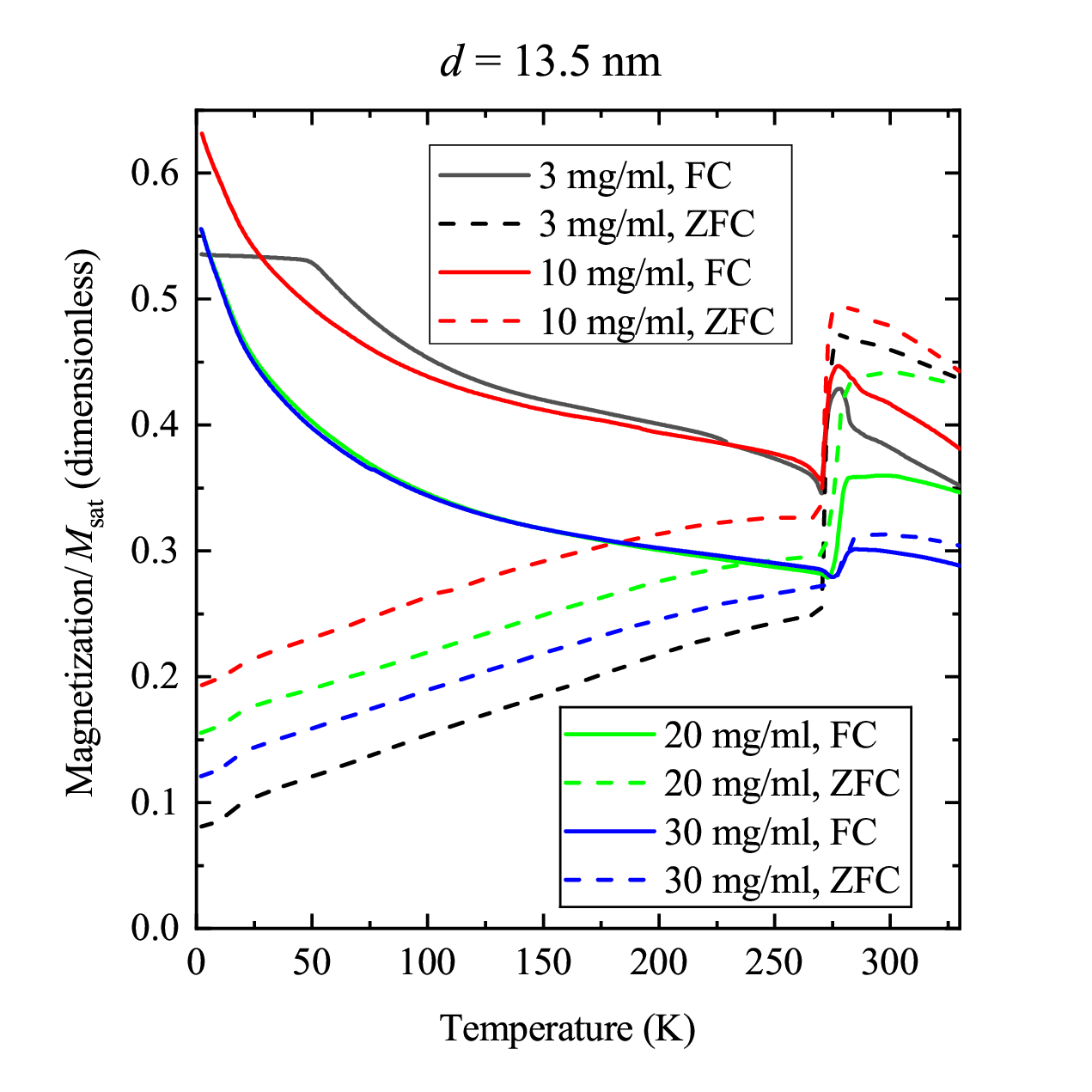} 
    \caption{Comparison of field-cooled (FC) and zero-field cooled (ZFC) magnetization curves for the 13.5 nm sample for 4 different concentrations after normalization with the saturated magnetization.}
    \label{SM_13x5_FC_ZFC}
\end{figure}

In Figure \ref{SM_13x5_FC_ZFC}. we show the field-cooled (FC) and zero-field cooled (ZFC) magnetization curves for the 13.5 nm sample for 4 different concentrations: 3, 10, 20, and 30 mg/mL. The data were normalized by the value of the saturated magnetization. 

\begin{figure}[h!]
    \includegraphics[width=0.5\linewidth]{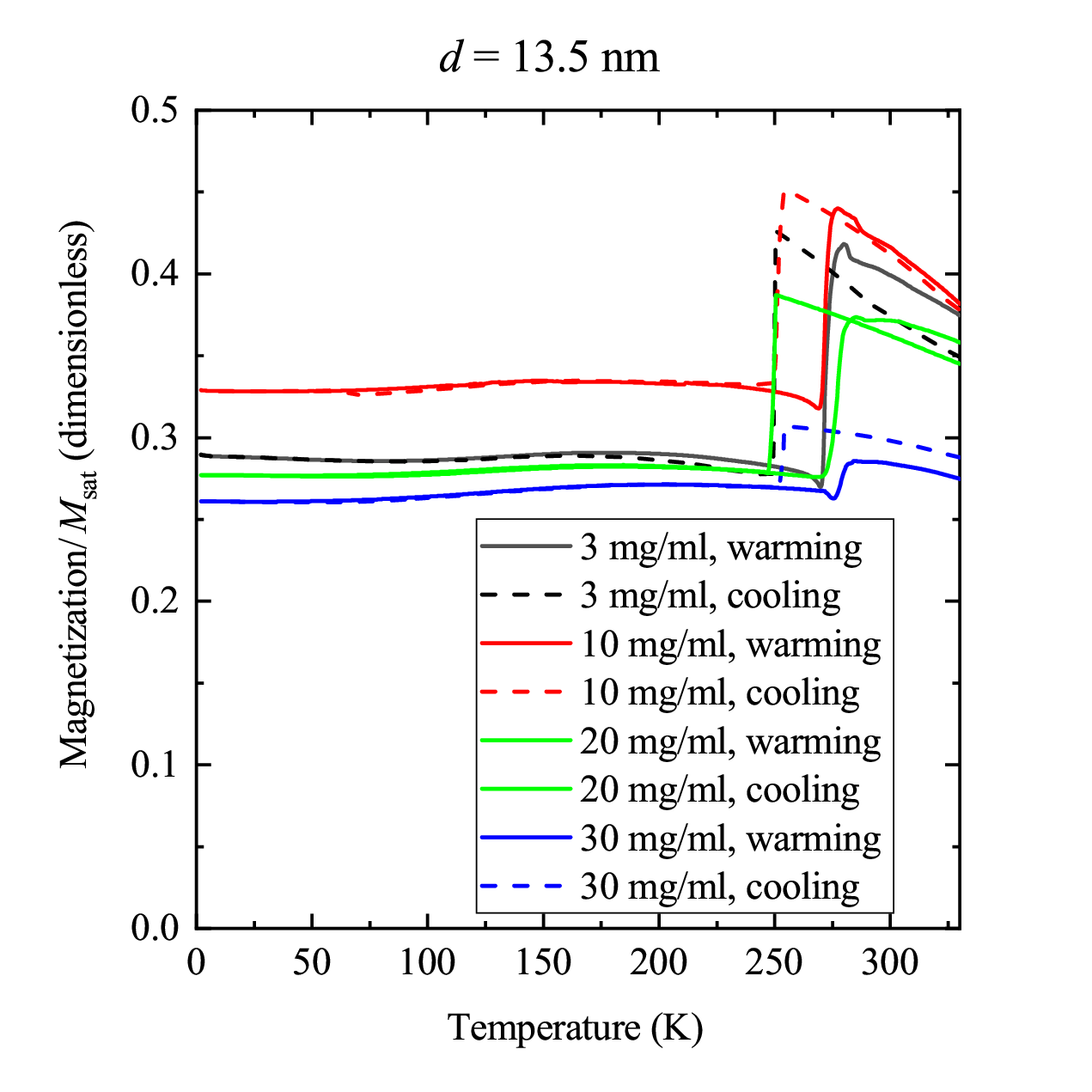} 
    \caption{Temperature-dependent thermal hysteresis curves for the 13.5 nm sample for varying concentrations. The measurement was performed at 100 Oe while cooling and warming the sample. The original Fe mass-normalized magnetization data were further normalized by the value of the saturated magnetization due to an error in the Fe amount estimate. This puts all curves onto similar values. Note that for the highest concentration, the thermal hysteresis curves markedly differ from the lower 3, 10, and 20 mg/mL values.}
    \label{SM_13x5_thermal_hysteresis}
\end{figure}

In Figure \ref{SM_13x5_thermal_hysteresis}. we show the thermal hysteresis curves for 4 concentrations: 3, 10, 20, and 30 mg/mL. Data were normalized by the value of the saturated magnetization. 

\newpage
\subsection{Systematic particle size-dependence}

\begin{figure}[htbp]
    \includegraphics[width=0.5\linewidth]{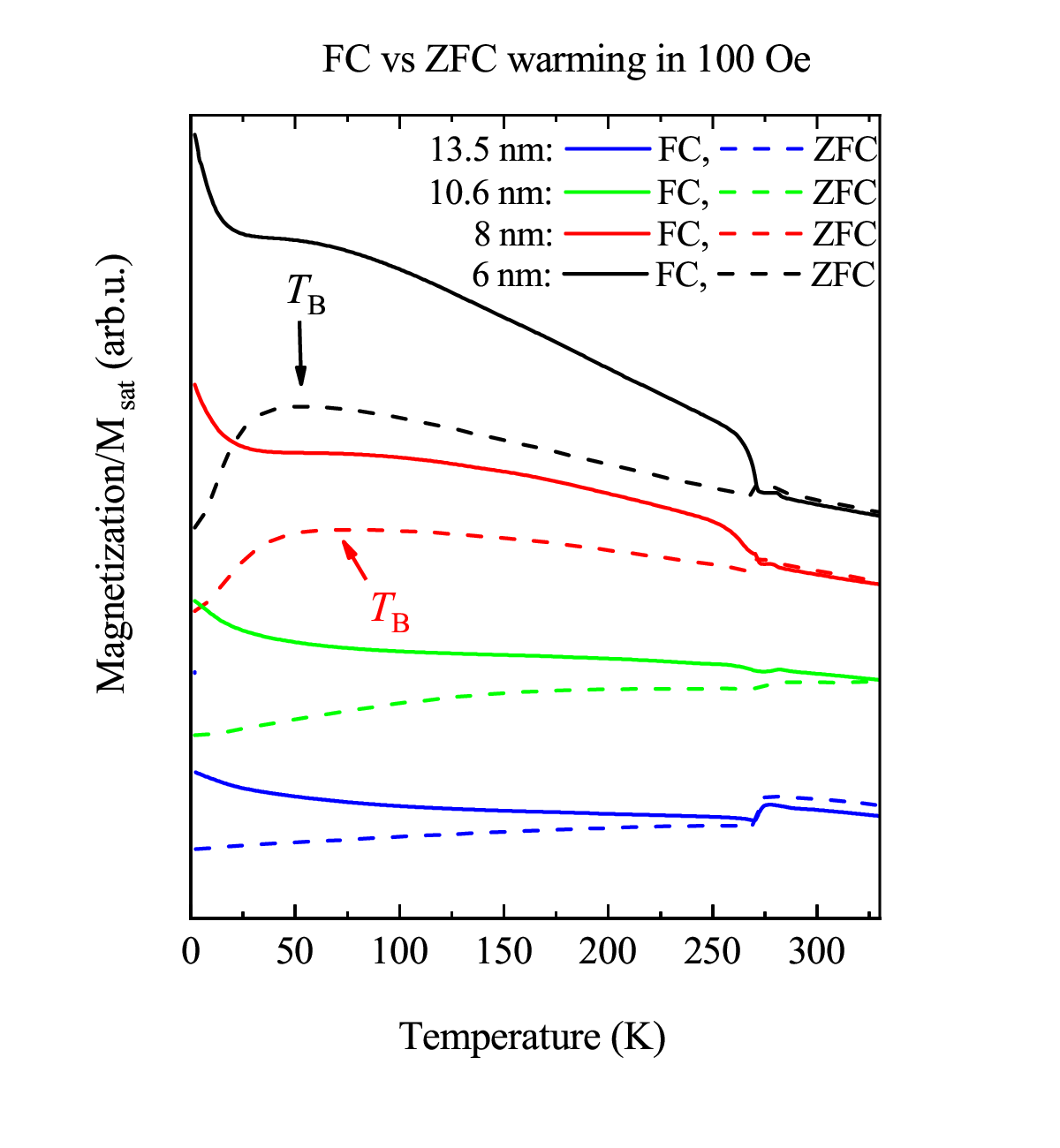} 
    \caption{Particle size dependence of the FC vs ZFC experiments (measured at 100 Oe) for the 6, 8, 10.6, and 13.5-nm diameter samples for the 10 mg/mL concentration. The measured magnetization values are normalized by the saturated magnetization value at 330 K and are offset for clarity. Arrows indicate the blocking temperature for the samples which show the Néel behavior.}
    \label{SM_FC_ZFC_6_8_10x6_13x5nm_10mgml_100Oe}
\end{figure}

Figure \ref{SM_FC_ZFC_6_8_10x6_13x5nm_10mgml_100Oe} shows the FC vs. ZFC measurements for the different particle sizes. The observations that are reported in the main text are reproduced with a systematic particle size dependence: the FC data drops at the water freezing point for the small diameter samples and it is continuous for the ZFC measurements. However, it increases for both the FC and ZFC measurements for the larger-diameter sample, which is a benchmark of the Brownian relaxation. The small-diameter samples are characterized by a blocking behavior at low temperatures (blocking temperatures are indicated by arrows), whereas the temperature-dependent magnetization is "flat" for the larger-diameter samples.

\begin{figure}[htbp]
    \includegraphics[width=0.5\linewidth]{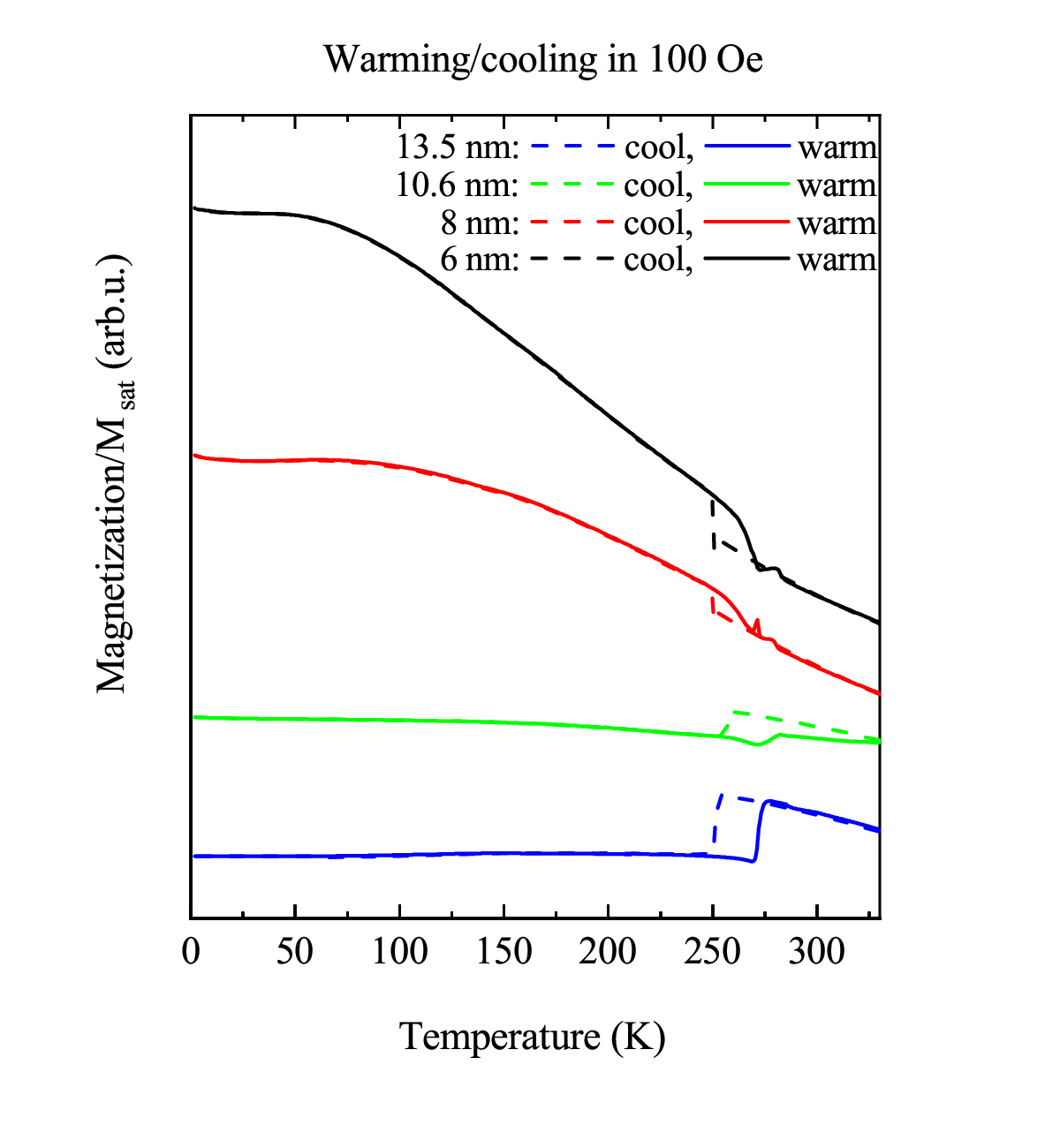} 
    \caption{Particle size dependence of the thermal hysteresis effect for the 6, 8, 10.6, and 13.5 nm diameter samples for the 10 mg/mL concentration. The measured magnetization values are normalized by the saturated magnetization value at 330 K and are offset for clarity.}
    \label{SM_Therm_hyst_6_8_10x6_13x5nm_10mgml_100Oe}
\end{figure}

Figure \ref{SM_Therm_hyst_6_8_10x6_13x5nm_10mgml_100Oe} shows the thermal hysteresis effect for the varying sample diameters of 6, 8, 10.6, and 13.5-nm samples at the 10 mg/mL concentration. The diameter dependence again shows several systematic diameter-dependent effects: 
the small diameter samples are characterized by a blocking behavior at low temperatures, whereas the temperature-dependent magnetization is "flat" for the larger diameter samples. At the water freezing point, the cooling magnetization data is above the warming for the small-diameter samples whereas this behavior is reversed for the larger-diameter ones. Both observations occur for the intermediate sample diameters, too but with reduced effect compared to the two extremal behavior at the 6 nm and 13.5 nm samples.

\newpage
\subsection{Results of Differential Scanning Calorimetry}

\begin{figure}[htbp]
    \centering
    \begin{subfigure}{0.49\linewidth}
        \centering
        \includegraphics[width=\linewidth]{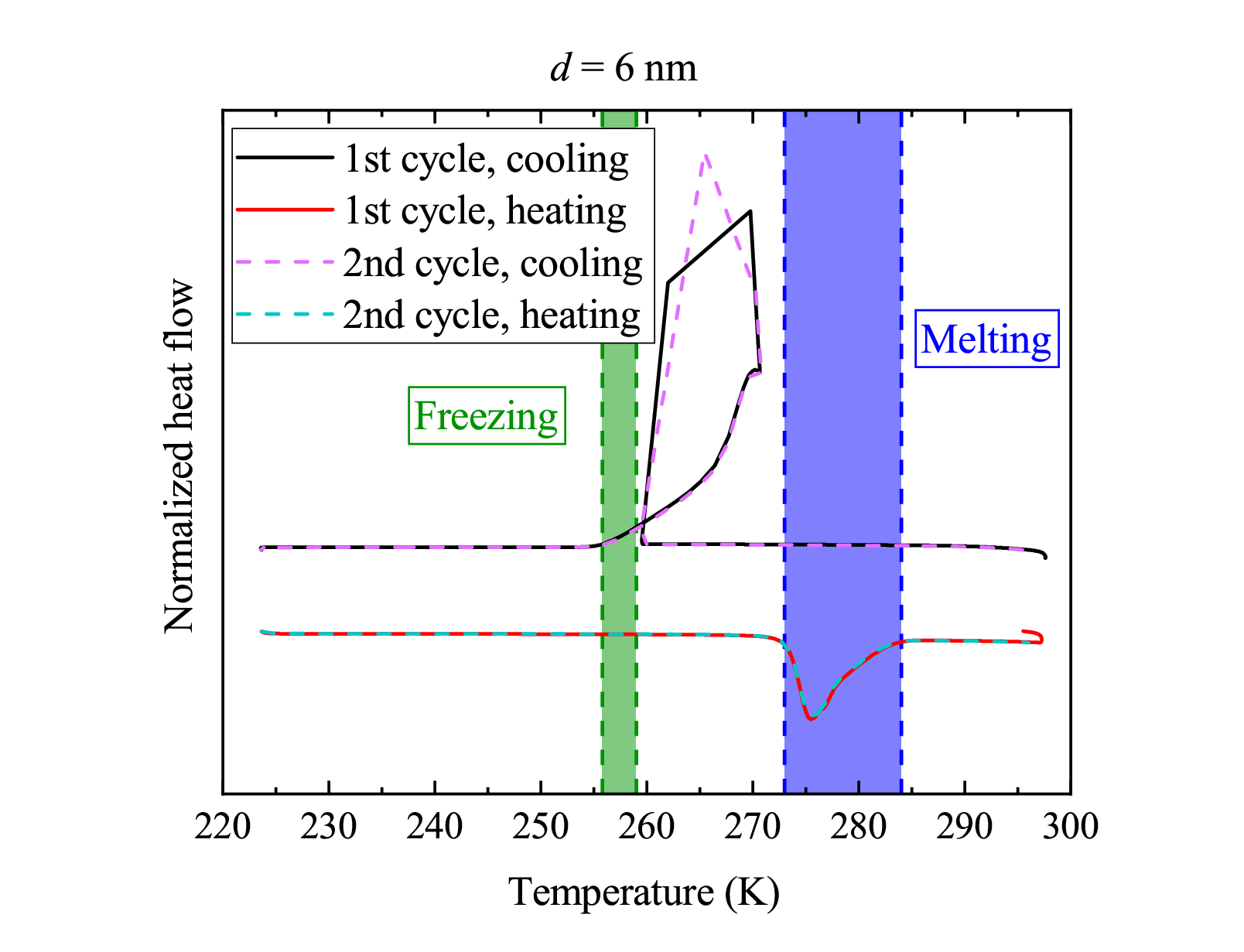} 
   \end{subfigure}
   \hfill
    \begin{subfigure}{0.49\linewidth}
        \centering
        \includegraphics[width=\linewidth]{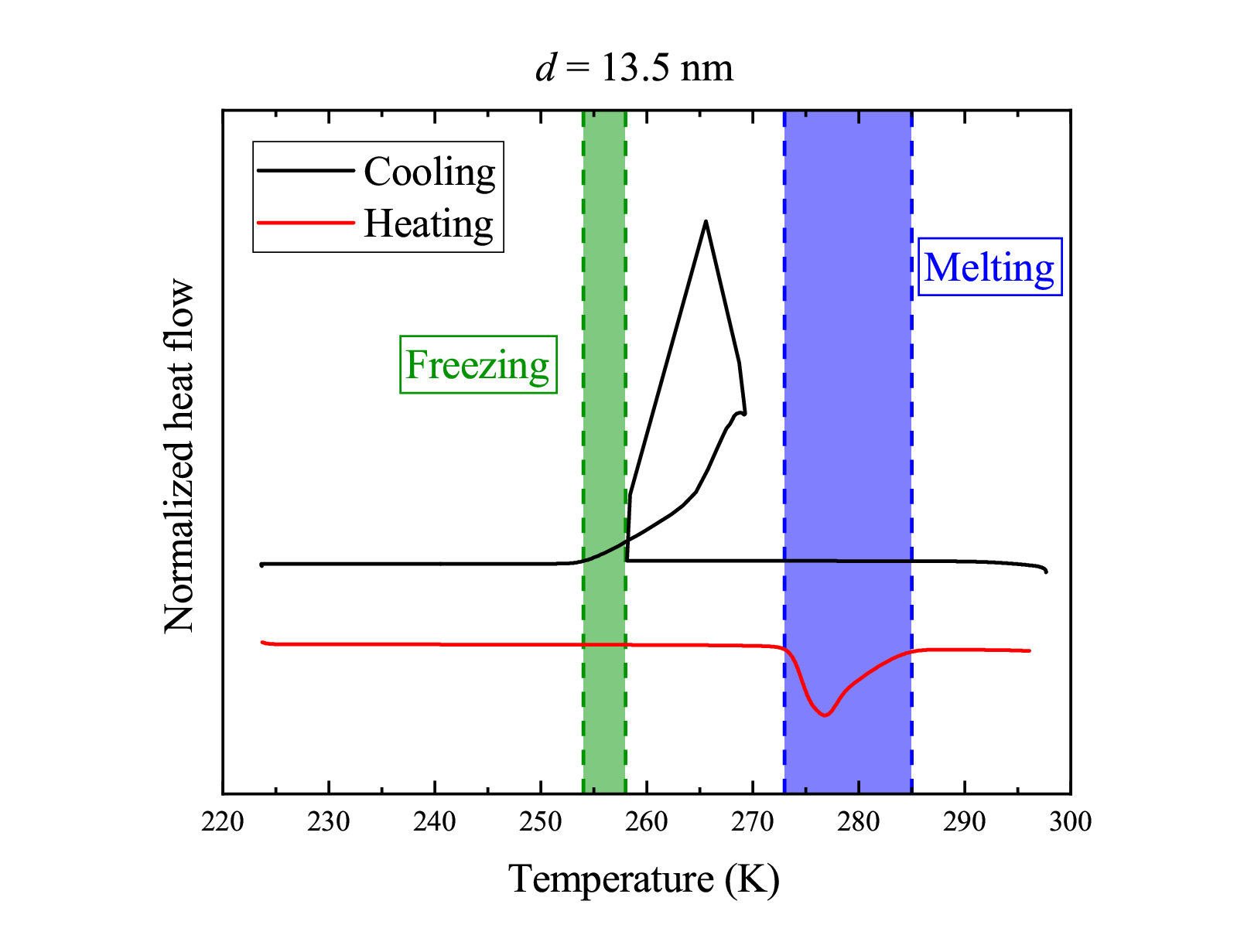} 
        \end{subfigure}
    \caption{Normalized heat flow as a function of temperature, curves are arbitrarily shifted for better visibility. Freezing and melting temperature ranges are denoted as shaded areas.}
    \label{DSC}
\end{figure}

When performing Differential Scanning Calorimetry measurements, the heat flow is recorded, which is needed for keeping a slow, uniform cooling of the system. Data were taken with a ramp rate of 5 K/min, a slower rate did not influence the measurement. The heat flow value was normalized by the sample mass depicted as a function of temperature (Figure \ref{DSC}).

Transition temperature ranges depicted in the main text were determined the following way: in the DSC data shown in Figure \ref{DSC} the melting and freezing transitions appear as peaks. The starts and ends of these peaks can be approximated as linear, and the transition points are determined as where these lines intersect the baseline. These melting and freezing temperature ranges are shown in the main text and more details are provided in the Supporting Materials.

To examine the effect of evaporation on the heat flow data, the measurement cycle was repeated for the 6-nm sample (Figure \ref{DSC} top). There is no visible change in the shape of the measured curves, except in the freezing peak, but that is clearly undersampled, so the deviation probably stems only from this. Its start and end, which are relevant for determining the transition points, however, do not suffer from this problem.

\end{document}